\documentclass[11pt]{article}
\usepackage{jcappub,natbib,float,caption,bm}

\usepackage{amssymb}
\usepackage{amsmath}
\usepackage{hyperref}

\usepackage{graphicx}
\usepackage{graphics}
\usepackage{dcolumn}
\usepackage{color}
\usepackage{rotate}
\usepackage{fancyhdr} 
\usepackage{comment}

\def\be{\begin{equation}}
\def\ee{\end{equation}}
\def\bea{\begin{eqnarray}}
\def\eea{\end{eqnarray}}
\def\nn{\nonumber}
\def\ba#1\ea{\begin{align}#1\end{align}}
\def\bcom#1\ecom{}
\newcommand{\vs}{\nonumber\\}
\def\rhobt{\tilde{\bar\rho}}
\newcommand{\Om}{\Omega_m}
\newcommand{\Omn}{\Omega_{m0}}

\def\VEV#1{\left\langle #1 \right\rangle}

\def\coH{\mathcal{H}}
\def\v#1{\bm{#1}}
\def\comment#1{}

\newcommand{\refeq}[1]{Eq.~(\ref{eq:#1})}          
\newcommand{\refeqs}[2]{Eqs.~(\ref{eq:#1})--(\ref{eq:#2})}          
\newcommand{\reffig}[1]{Fig.~\ref{fig:#1}}          
\newcommand{\refsec}[1]{Sec.~\ref{sec:#1}}
\newcommand{\refapp}[1]{App.~\ref{app:#1}}

\newcommand{\bfx}{\mathbf{x}}
\newcommand{\bfk}{\mathbf{k}}

\def\rhob{\bar\rho}

\title{On Separate Universes}

\author[a]{Liang Dai,}

\author[b]{Enrico Pajer,}

\author[c]{Fabian Schmidt}

\affiliation[a]{Department of Physics and Astronomy, Johns Hopkins University, 3400 N. Charles St., Baltimore, MD 21218, USA}

\affiliation[b]{Institute for Theoretical Physics and Center for Extreme Matter and Emergent Phenomena,
Utrecht University, Leuvenlaan 4, 3584 CE Utrecht, The Netherlands}

\affiliation[c]{Max-Planck-Institut f\"ur Astrophysik, Karl-Schwarzschild-Str. 1, 85741 Garching, Germany}

\abstract{The \textit{separate universe conjecture} states that in General Relativity a density
perturbation behaves locally (i.e.~on scales much smaller than the
wavelength of the mode) as a separate universe with different background
density and curvature.   We prove this conjecture for a spherical compensated tophat
density perturbation of arbitrary amplitude and radius in $\Lambda$CDM.  
We then use
\emph{Conformal Fermi Coordinates} to generalize
this result to scalar perturbations of arbitrary configuration and scale
in a general cosmology with a mixture of fluids,
but to linear order in perturbations.  In this case, the separate
universe conjecture holds for the isotropic part of the perturbations.  
The anisotropic part on the other hand is exactly
captured by a tidal field in the Newtonian form. We show that the separate universe picture is restricted to scales larger than
the sound horizons of all fluid components. We then derive an expression
for the locally measured \textit{matter bispectrum} induced by a long-wavelength mode of arbitrary wavelength, a new result which in standard perturbation theory is equivalent to a relativistic second-order calculation. 
We show that nonlinear gravitational dynamics does \textit{not} generate observable contributions that scale like local-type non-Gaussianity $f^{\rm loc}_{\rm NL}$, and hence does {\it not} contribute to a scale-dependent galaxy bias $\Delta b \propto k^{-2}$ on large scales; rather, the locally measurable long-short mode coupling assumes a form essentially identical to subhorizon perturbation theory results, once the long-mode density perturbation is replaced by the synchronous-comoving gauge density perturbation.  
Apparent $f^{\rm loc}_{\rm NL}$-type contributions arise through projection effects on photon propagation, which depend on the specific large-scale structure tracer and observable considered, and are in principle distinguishable from the local mode coupling induced by gravity. We conclude that any observation of $f^{\rm loc}_{\rm NL}$ beyond these projection effects signals a departure from standard single-clock inflation.
}
\begin{document}

\maketitle
\flushbottom

%%%%%%%%%%%%%%%%%%%%%%%%%%%%%%%%%%%%%%%%%%%%%%%%%%%%%%%%%%%%%%%%
\section{Introduction and summary of results}
\label{sec:intro}
%%%%%%%%%%%%%%%%%%%%%%%%%%%%%%%%%%%%%%%%%%%%%%%%%%%%%%%%%%%%%%%%

An accurate theoretical description of the nonlinear large-scale structure is one of the major goals of theoretical cosmology.  Although it is by nature a difficult problem, nonlinear structure formation is a key ingredient toward decoding the initial cosmological perturbations in the primordial Universe and testing the large-scale dynamics of gravitation.  This is especially important given the plethora of current or incoming large-scale structure surveys aiming at measurements of unprecedented accuracy. A general relativistic treatment is essential if perturbations on scales close or comparable to the horizon are being considered.  Since the initial conditions of structure (e.g., due to inflation) are generally specified when they are outside the horizon, this issue necessarily arises when attempting to connect structure in the observable universe to the initial conditions.  

One of the primary reasons of why we would like to connect late-time observables of large-scale structure to the initial conditions at nonlinear order is the study of interactions during inflation, in particular through~\textit{primordial non-Gaussianity}.  Specifically, the interaction between long-wavelength and short-wavelength perturbations 
is an interesting diagnostic of inflationary physics, as it can distinguish between one or more light degrees of freedom during inflation.  This
can be most directly measured from the squeezed configuration of the three-point (bispectrum) or higher-point functions, where ``squeezed'' means that one Fourier mode has a wavelength much longer than all the other modes.  For single-clock inflation, model-independent squeezed-limit consistency relations are known to hold~\cite{Maldacena:2002vr,Creminelli:2004yq,Kehagias:2013paa,Cheung:2007sv,Creminelli:2012ed,
Hinterbichler:2012nm,Senatore:2012wy,Hinterbichler:2013dpa,Goldberger:2013rsa}, the simplest representative of which reads $f^{\rm loc}_{\rm NL} = 5/12 (n_s-1)$, where $f^{\rm loc}_{\rm NL}$ is the amplitude of local-type non-Gaussianity in the primordial (Newtonian gauge) potential. 
Physically, they encode the existence of a single preferred clock.  A detection of local-type primordial non-Gaussianity that departs from the unique single-field prediction would prove the existence of more than one light field during inflation. 

In terms of the matter three-point function, local-type non-Gaussianity schematically leads to a squeezed limit of~\footnote{A prime on the bispectrum indicates that a prefactor of $(2\pi)^3 \delta_D(\bfk_1 + \bfk_2 + \bfk_L)$ imposed by statsitical homogeneity has been dropped. We will use this notation in the rest of the paper.  For the moment we will disregard the issue of which gauge $\delta(\bfk)$ is to be evaluated in.}
\bea
\hspace{-0.5cm}
\VEV{\delta(\bfk_1) \delta(\bfk_2) \delta(\bfk_L)}{}' & \propto & f^{\rm loc}_{\rm NL}\, \left( a H / k_S \right)^2\,P_\delta(k_L)\,P_\delta(k_S), \quad {\rm where} \,\, \bfk_S = (\bfk_1-\bfk_2)/2\,,
\label{eq:BfNL}
\eea
where corrections are suppressed by $(k_L/k_S)^2$ and we have assumed $k_L < k_{\rm eq}$ so that the transfer function is unity. While consistency relations are a statement about transformation properties under diffeomorphisms, 
in a local observer's frame the diffeomorphism freedom is completely fixed by the observer's proper time and distance standards.  As shown in \cite{Pajer:2013ana} (see also \cite{Creminelli:2004pv,Bartolo:2011wb,Creminelli:2011sq,Dai:2013kra}), this implies that squeezed-limit consistency relations drop out of locally measurable quantities.  
Non-single-field models which violate the consistency relation will on the other hand contribute to physical observables and can be measured in LSS surveys
through the bispectrum or scale-dependent bias $b(k) \propto k^{-2}$~\cite{Dalal:2007cu}.

Thus, an important question to ask is whether nonlinear gravitational evolution of the matter density field, neglected in \refeq{BfNL}, will lead to contributions of $f^{\rm loc}_{\rm NL}$-type to matter statistics. Investigating this requires a relativistic treatment of perturbations at second order since both the long- and short-wavelength modes have to be followed through horizon crossing.  This has been done for example in \cite{Bartolo:2005xa,Verde:2009hy,Noh:2003yg}. Unfortunately, standard higher-order perturbative approaches for nonlinear mode-coupling are often plagued by unphysical gauge artifacts due to the freedom to parameterize perturbations in different coordinate systems (e.g., \cite{Naruko:2013aaa}). 

An alternative approach to this problem is based on the intuition that a local observer who has access only to short comoving distances $\sim 1/k_S$, which are much smaller than the scale of variation of the long-wavelength modes, would interpret local small-scale physics within a Friedmann-Lema\^itre-Robertson-Walker (FLRW) background, i.e. a ``separate universe''~\cite{Lamaitre:1933}. To formulate this intuition rigorously, one starts with a coarse-grained universe in which small-scale inhomogeneities with $k_S$  greater than a fixed comoving scale $\Lambda$ are first smoothed out. One then studies how small-scale structure with $k_S > \Lambda$ evolves in this ``local background'' spacetime, instead of the ``global'' background universe. To solve small-scale clustering in this modified background, one can take a Lagrangian perspective --- if non-gravitational forces are negligible at large distances, one can follow inertial observers in the coarse-grained universe and their vicinity. As initially proposed in \cite{Pajer:2013ana} and rigorously proven in \cite{Dai:2015rda}, one can locally construct a frame, \emph{Conformal Fermi Coordinates} (CFC), valid across a comoving distance of order $\Lambda^{-1}$ at all times, which realizes this physical picture.  In CFC, the observer interprets the small-scale structure around her as evolving in an FLRW universe that is modified due to long-wavelength modes with comoving wavenumber $k_L < \Lambda$, with corrections quadratic in the spatial distance to the observer
\bea
\label{eq:CFC-metric-form}
ds^2 & = & a^2_F(\tau_F) \left[ -d\tau_F^2 + (1+K_F\, \bfx^2_F/4)^{-2}\delta_{ij}\, dx^i_F dx^j_F + \mathcal{O}[(x^i_F)^2] \right]\,,
\eea
where $a_F$ is the local scale factor and $K_F$ is the local spatial curvature.  
The fact that the metric is valid over a fixed \emph{comoving} scale is
very important.  This means that we can apply the CFC frame even when
the small-scale modes are outside the horizon.  This is crucial in order
to connect to the initial conditions given by inflation.  
Therefore, within CFC one can keep track of any general relativistic effect that could arise as either long- or short-wavelength perturbations cross the horizon.   

This ``separate universe'' approach enjoys at least two advantages. On the one hand, gauge freedom is eliminated because the calculation directly yields what the CFC observer physically measures. On the other hand, it easily relates to small-scale structure formation in homogeneous FLRW backgrounds, which we know very well how to handle. An important requirement for this is that $a_F(\tau_F)$ 
indeed corresponds to a locally observable scale factor. This can be
achieved easily by using the local velocity divergence as described in \cite{Dai:2015rda} and \refsec{CFNC-main} below.  

The separate universe picture strictly holds if the local spacetime is 
indistinguishable from an unperturbed FLRW universe within a small region of space but for an extended (or even infinite) duration of time.  We find that
this is in general only true if the observer and all components of matter with equation of state $w \neq -1$ are comoving.  This implies that the separate universe picture is restricted to scales larger than the sound horizon of all relevant fluids.

One of the main results of this work is the precise form
of the CFC metric \refeq{CFC-metric-form} and the relation of the 
local scale factor $a_F$ and curvature $K_F$ to the long-wavelength
perturbations, which we assume to be linear.  
The latter relation becomes simplest in synchronous-comoving (sc) gauge,
where we obtain (for a flat global cosmology) the surprisingly concise relations
\be
\frac{a_F(\tau_F)}{a(\tau_F)} = 1 - \frac13 \Delta_{\rm sc}(\tau_F)
\quad\mbox{and}\quad
\frac{K_F}{H_0^2} = \frac53 \Omega_{m0} \frac{\Delta_{\rm sc}(\tau)}{D(\tau)}\,,
\label{eq:KFsc2}
\ee
where $\Delta_{\rm sc}$ is the long-wavelength density perturbation, 
$a(\tau)$ is the scale factor of the background universe, 
and $D(\tau)$ is the linear growth factor normalized to $a(\tau)$ during matter domination. 
Furthermore, the tidal corrections $\mathcal{O}[(x^i_F)^2]$ in \refeq{CFC-metric-form} can be 
written as
\ba
ds^2 = a_F^2(\tau_F) \Bigg\{ & - \left[ 1 + 2 \Phi_F \right] d\tau_F^2 
+ \left[ 1 - 2\Psi_F \right] \frac{\delta_{ij} dx_F^i dx_F^j}{\left(1+K_F\, \bfx_F^2/4\right)^2} \Bigg\}\,.
\ea
where $\Phi_F,\,\Psi_F$ are simply related to the scalar potentials in the conformal Newtonian gauge through
\ba\label{eq:ee}
\Phi_F \equiv \frac12 \left( \partial_k \partial_l \Phi - \frac13 \delta_{kl} \partial^2 \Phi \right) x_F^k x_F^l\,,
\ea
and an analogous relation holds for $\Psi_F$ in terms of $\Psi$.  
Note that $\Phi_F,\,\Psi_F$
involve the trace-free part of the second spatial derivatives of $\Phi$ and $\Psi$, i.e. the Newtonian tidal tensor,
and thus vanish for an isotropic perturbation.

These results can be summarized in words as:
\paragraph{}\emph{ On scales larger than the sound horizon of the fluid, the effect of a long-wavelength mode as measured locally is completely captured by a modified local scale factor and spatial curvature, \refeq{KFsc2}, and a pure tidal field, \refeq{ee}.}

\medskip

Ref.~\cite{Pajer:2013ana} showed that
the consistency relation for single-clock inflation, transformed into CFC,
states that there is no primordial correlation between long- and short-wavelength
potential perturbations. This provides the trivial initial conditions for the small-scale fluctuations in CFC.  Integrating the growth of small-scale density perturbations $\delta$ in the presence of a long-wavelength mode $\Delta(\bfk_L)$, we then obtain the leading contribution from nonlinear gravitational evolution, valid for arbitrarily small $\bfk_L$,
\bea
\label{eq:d2intro}
\delta_E & = & \delta^{(1)} + \frac{13}{21} \Delta_{\rm sc}\, \delta^{(1)} + \frac4{7}\, K^\Delta_{ij} \, \left( \frac{\partial^i\partial^j}{\partial^2} \delta^{(1)} \right)  + K^\Delta_{ij}\,x^i \partial^j \delta^{(1)}\,.
\eea
Here, $\delta^{(1)}$ is the linear small-scale density field (i.e.~in the absence of long-wavelength modes), while $\Delta_{\rm sc}$ is the long-wavelength density perturbation (in synchronous gauge), and $K^\Delta_{ij} = (\partial_i\partial_j/\partial^2 - \delta_{ij}/3) \Delta_{\rm sc}$ is proportional to the tidal tensor.  
\refeq{d2intro} shows that the \textit{locally
observable small-scale density in single-clock inflation
has no contributions that scale as local-type non-Gaussianity} (this holds not just for quadratic $f_{\rm NL}^{\rm loc}$ but equally for higher order terms $g_{\rm NL}^{\rm loc}, ...$).  
This implies that there is no scale-dependent bias of large-scale structure tracers even when taking into account nonlinear gravitational evolution fully relativistically.  
We further derive the leading squeezed-limit matter bispectrum from \refeq{d2intro} [\refeq{squeezed-local-matter-bispec}] 
which proves this point. Specifically, this is the bispectrum which would be seen by a ``central observer'' if synchronous observers distributed on 
the past lightcone of the central observer communicated their local densities and power spectra of small-scale fluctuations.  
This bispectrum is suppressed over local-type non-Gaussianity by a factor of $k_L^2/\coH^2$.  

In order to provide predictions for correlation functions measured from Earth, we need to include photon propagation (``projection'') effects.  These correspond to mapping the locally measurable observables in physical space to the observer's coordinates of measured photon redshift and arrival direction. These effects do contain contributions that correspond to order
unity $f^{\rm loc}_{\rm NL}$.  However, we stress that \textit{projection effects are purely kinematical, not dynamical}, and depend on the details of the
tracer considered.  Thus, they are in principle distinguishable from 
the locally measurable local-type non-Gaussianity, which, as we have shown,
is not generated by gravitational evolution and must come from the initial conditions.  
A detailed treatment of projection effects is beyond the scope of
this paper.  For recent literature on this subject, see \cite{Pajer:2013ana,Kehagias:2015tda,Bertacca:2014wga,DiDio:2014lka}.

The remainder of this paper is organized as follows. In \refsec{proof}, we demonstrate the concept of ``separate universe'' by presenting a full general relativistic proof that a compensated spherical top-hat region of overdensity embedded in a FLRW universe behaves exactly as a separate FLRW universe (with different spatial curvature).  In \refsec{CFNC-main} we specialize to the CFC for a free-falling observer in the presence of a long-wavelength scalar metric perturbation at linear order. We discuss in \refsec{local-friedmann-eqs} the sufficient and necessary conditions for which local FLRW expansion is exact, i.e. for which the spacetime is truly locally indistinguishable from an FLRW universe.  The explicit calculation of the locally measurable squeezed matter bispectrum for an Einstein-de Sitter (EdS) universe ($\Omega_m=1$) with single-clock initial conditions is presented in \refsec{app-squeezed-matter-bispec}. Concluding remarks are given in \refsec{concl}.

Regarding our notation, up to including \refsec{CFNC-main} we distinguish between the global comoving position $x^i$ and the CFC comoving position $x^i_F$, so as to highlight the distinction between the global coordinate system and the CFC.  After that we will exclusively use spatial CFC coordinates and denote them simply by $x^i$. On the other hand, we always explicitly distinguish between the CFC times $\tau_F\,, t_F$ and the global times $\tau\,,t$.

%%%%%%%%%%%%%%%%%%%%%%%%%%%%%%%%%%%%%%%%%%%%%%%%%%%%%%%%%%%%%%%%%%%%%%%%%%%
%%%%%%%%%%%%%%%%%%%%%%%%%%%%%%%%%%%%%%%%%%%%%%%%%%%%%%%%%%%%%%%%%%%%%%%%%%%
\section{Proof of the separate universe conjecture for a compensated tophat}
\label{sec:proof}

\begin{figure}[t!]
\centering
\includegraphics[width=0.49\textwidth]{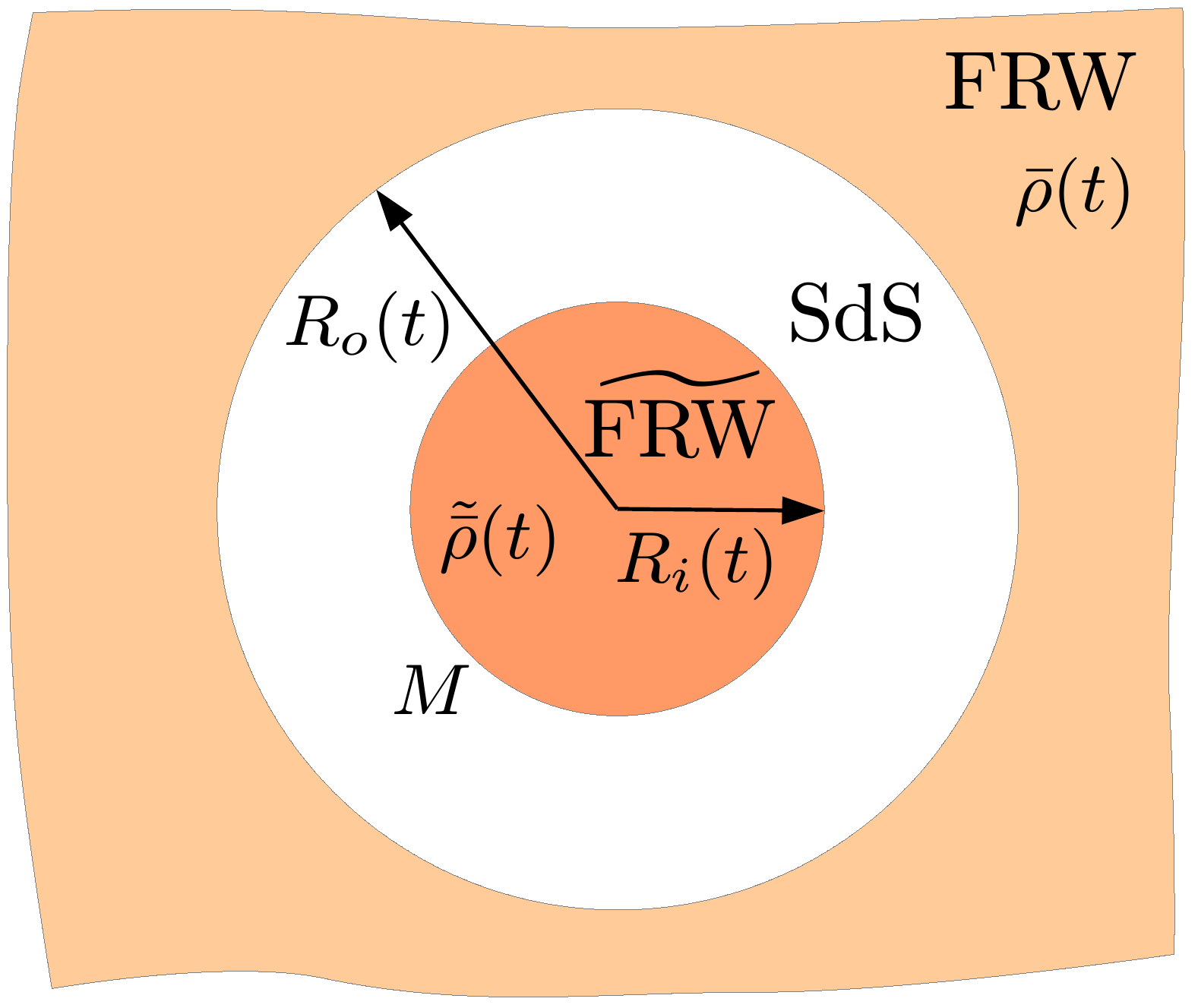}
\caption{Illustration of the setup used to prove the separate universe
approach. 
\label{fig:sketch}}
\end{figure}

The ``separate universe conjecture'' states that a spherically symmetric
perturbation in an FLRW background (taken to be $\Lambda$CDM in this section) 
behaves like a separate FLRW universe
with different matter density and curvature.  This holds up to higher
spatial derivative corrections, but at all times.  
We now provide a proof of this statement for a spherical compensated tophat, or just tophat for short (see figure \ref{fig:sketch}), without making any assumptions
about the amplitude or wavelength of the perturbation (i.e.~we do not
assume that it is subhorizon).  Higher derivative corrections are avoided
by assuming a homogeneous density perturbation.  

Consider a ``background'' FLRW universe described in comoving spherical coordinates
$(t, r_o, \theta,\phi)$ by 
\be
ds^2_{\rm FLRW} = -dt^2 + a^2(t) \left[ (1-K r_o^2)^{-1} dr_o^2 + r_o^2\, d\Omega^2 \right]\,,
\label{eq:dsFLRW}
\ee
where we have allowed for spatial curvature $K$. The label ``$o$'' stands for ``outside'', as it will become clear shortly. Note that $a(t) r_o$ is the
\emph{area radius}, so that the proper surface area of a 2-sphere of radius
$r_o$ is $4\pi a^2(t) r_o^2$.\footnote{The geodesic radius on the other hand, i.e.~the proper radius along a radial geodesic, is given by
$\chi$, where $\chi$ is defined through $r_o = \sin_K ( \sqrt{|K|} \chi )/\sqrt{|K|}$ and $\sin_K x= \sin x,\,x,\,\sinh x$ for $K >0,\,=0,\,<0$, respectively.}  This will become very useful in the following. We assume that the universe is filled with dust of uniform density $\rhob(t)$ and a cosmological constant.
Now consider cutting out a sphere of comoving radius $r_o = R_{o,c}$ from the FLRW background (where ``$c$'' stands for ``comoving''), and 
collapsing the matter in the sphere into a point mass $M$.  
If we choose $R_{o,c}$ to be constant in time, then $M$ is conserved.\footnote{In a flat background, this mass is simply $(4\pi/3)\,\rhob(t) a^3(t) R_{o,c}^3$, while for $K\neq 0$, the mass-radius relation is more complicated.  Independently of curvature, the interior mass scales as $a^3(t)$ for constant $R_{o,c}$ and is conserved.} 
The authors of \cite{Einstein/Straus} first showed for a matter-dominated universe that there is a unique
solution consisting of a Schwarzschild metric interior to $R_{o,c}$ smoothly
matched to the FLRW background \refeq{dsFLRW} at $r_o = R_{o,c}$ for all times $t$ (more precisely, the metric is continuously differentiable at the boundary).  This
result has since been generalized to include a cosmological constant
\cite{Balbinot:1988zc,carrera/giulini}, in which case the point mass
metric becomes the Schwarzschild-de Sitter solution:
\ba
ds^2_{\rm SdS} =\:& - V(r_s) dt_s^2 + V^{-1}(r_s) dr_s^2 + r_s^2 d\Omega^2 \vs
V(r_s) =\:& 1 -\frac{2GM}{r_s} - \frac13 \Lambda r_s^2\,,
\label{eq:dsSdS}
\ea
where the label ``$s$'' stands for ``Schwarzschild''. Hence, the matter surrounding the shell is oblivious to the fact that the interior has
been collapsed to a black hole, while a test particle anywhere inside $ R_{o,c} $ is oblivious to the surrounding homogeneous matter.  Note that
by matching the angular part of the metric, we can immediately identify
$r_s = a(t) r_o$.  Moreover, $r_s$ is the area radius for $ds^2_{\rm SdS}$.  
In the interior coordinates, the boundary is thus set by $r_{s,\rm bound} =  R_o(t)$.  Using this relation, it is then easy to show 
that at the boundary the geodesic equation derived from the
metric \refeq{dsSdS} exactly matches that given by the FLRW background
\refeq{dsFLRW}, which is simply $R_o(t)/a(t) = R_{o,c} = $~const, once the Friedmann equation
for $a(t)$ is inserted.  This is of course a consequence of the metric being
continuously differentiable at the boundary.

Consider now the case where we do not collapse the matter inside $R_o$
into a black hole, but compress it to a finite radius $r_s = R_i(t)$ (where ``$i$'' stands for ``inside''), 
maintaining
a homogeneous density $\rhobt$ (see \reffig{sketch}).  
Birkhoff's theorem \cite{Birkhoff}, generalized to include a cosmological
constant, states that the unique spherically symmetric vacuum solution to 
Einstein's equation is Schwarzschild-de Sitter.  Thus, \refeq{dsSdS}
still describes the spacetime outside of the mass, $r_s > R_i$.  We now
perform the exact same matching as derived by \cite{Einstein/Straus,Balbinot:1988zc,carrera/giulini}, but inverting outside and inside.  
In fact, nothing in the matching is particular to the case of vacuum inside
a homogeneous matter distribution; it equally applies to a homogeneous
density distribution inside vacuum.  Thus, there exists a unique FLRW
solution of the form \refeq{dsFLRW} with density $\rhobt$ that smoothly
matches to the Schwarzschild-de Sitter metric at $R_i$, where $\rhobt$
is determined by mass conservation.

%!!!!!!!!!!!!!!!!!!!!!!!!!!!!!!!!!!!!!!!!!!!
%!!!!!!!!!!!!!!!!!!!!!!!!!!!!!!!!!!!!!!!!!!!

Moreover, we can use the 
geodesic motion that follows from the metric \refeq{dsSdS} to derive
the scale factor $\tilde a(t)$ in the interior FLRW solution.  
The geodesic equation for \refeq{dsSdS} and purely radial motion,
$u^\mu \equiv dx_s^\mu/dt = (u^0, u^r)$ where $t$ is the proper time, becomes
\be
\frac{du^r}{dt} + \frac12 \frac{dV(r_{s}(t))}{dt} = 0\,,
\ee
which can be integrated to give
\be
\dot R_i^2(t) + V(R_i(t)) = {\rm const}\,,
\ee
where we used $\dot R_i \equiv dR_i/dt = u^r$.  Inserting \refeq{dsSdS},
we obtain
\ba
\dot R_i^2 =\:& \frac{2 G M}{R_i} + \frac13 \Lambda R_i^2 + {\rm const}\,.
\label{eq:dotRi}
\ea
This can be turned into a Friedmann equation by introducing a scale
factor $\tilde a(t) \propto R_i(t)$ different from the ``background'' $a(t)$.  
The normalization of the scale factor is arbitrary, and 
we can choose to normalize to $M = (4\pi/3) \rhobt(t) R_i^3(t_0) \tilde a^3(t)$ 
where $t_0$ is some reference time and $\rhobt(t) \propto \tilde a^{-3}(t)$.  
Dividing \refeq{dotRi} by $R_i^2$, we obtain
\ba
\tilde H^2 \equiv \left( \frac{\dot{\tilde a}}{\tilde a} \right)^{2} =\:& \frac{8\pi G}{3} \rhobt(\tilde a) + \frac13 \Lambda  - \frac{K_i}{\tilde a^2}\,.
\ea 
We have relabeled the integration constant suggestively as $-K_i$.  
This is indeed the Friedmann equation for a $\Lambda$CDM universe
with background matter density $\rhobt$ and curvature $K_i$.  Assuming that the perturbation
was initialized with a very small amplitude at an early time $t_i$ when the
effect of $\Lambda$ is negligible, one can then
subtract the background FLRW equation following from \refeq{dsFLRW}, and
linearize in $(R_i - R_o)/R_o$ at $t_i$.  
As shown in \refsec{local-friedmann-eqs} below (see also \cite{Wagner:2014aka}), this yields a relation between the initial
(linear) density perturbation and curvature
\be
K_i = \frac53 \Om H_0^2 a^{-1}(t_i) \frac{\rhobt(t_i)-\rhob(t_i)}{\rhob(t_i)}\,,
\label{eq:Ki}
\ee
where for simplicity we have assumed that the background FLRW universe is flat 
(although this is not a necessary assumption).  

Thus, we have proved that, at least for the case of a compensated tophat density profile,
a spherically symmetric perturbation in a $\Lambda$CDM background
\emph{evolves exactly as a separate curved} $\Lambda$CDM \emph{universe}.  We have
not assumed that the perturbation is small or that the scales $R_o,\,R_i$
are subhorizon.  Indeed, neglecting $\Lambda$ for simplicity, 
the Schwarzschild-de Sitter metric in the vacuum
surrounding the perturbation cannot be perturbatively approximated as
Minkowski if
\be
\frac{G M}{R_i} \sim 1\,, \quad\mbox{hence}\quad
(H_i R_i)^2 \sim 1\,,
\ee
since $G \rhobt \sim H_i^2$.  Thus, for a horizon-scale perturbation, the
vacuum exterior is far from Minkowski.  Nevertheless, the separate universe description is exact.  In fact, an observer in 
the vacuum region surrounding the overdensity would see this separate
universe as a black hole.  

One might wonder how this can be applied to an underdensity (void) rather
than an overdensity.  In this case, the inner FLRW solution which we
have called $\rhobt$ now becomes the background $\rhob$, while the outer,
less dense FLRW solution becomes the ``perturbation'' $\rhobt$.  From
the perspective of an observer inside either FLRW solution, the situation
is completely symmetric, as they cannot tell by any local measurement
whether they are embedded in a larger ``background'' universe.

In \refsec{local-friedmann-eqs}, we will return to the separate universe picture in the
context of general scalar perturbations in a cosmology with multiple fluids.

%%%%%%%%%%%%%%%%%%%%%%%%%%%%%%%%%%%%%%%%%%%%%%%%%%%%%%%%%%%%%%%%%%%%%%%%%%%%%%%%
%%%%%%%%%%%%%%%%%%%%%%%%%%%%%%%%%%%%%%%%%%%%%%%%%%%%%%%%%%%%%%%%%%%%%%%%%%%%%%%%
\section{Scalar perturbations in the CFC frame}
\label{sec:CFNC-main}
%%%%%%%%%%%%%%%%%%%%%%%%%%%%%%%%%%%%%%%%%%%%%%%%%%%%%%%%%%%%%%%%%%%%%%%%%%%%%%%%
%%%%%%%%%%%%%%%%%%%%%%%%%%%%%%%%%%%%%%%%%%%%%%%%%%%%%%%%%%%%%%%%%%%%%%%%%%%%%%%%

In this section, we describe how long-wavelength adiabatic scalar perturbations
are treated in the CFC frame.  This provides the basis for reconsidering
the separate universe ansatz in the following sections.

%%%%%%%%%%%%%%%%%%%%%%%%%%%%%%%%%%%%%%%%%%%%%%%%%%%%%%%%%%%%%%%%%%%%
\subsection{Review of Conformal Fermi Coordinates (CFC)}
%%%%%%%%%%%%%%%%%%%%%%%%%%%%%%%%%%%%%%%%%%%%%%%%%%%%%%%%%%%%%%%%%%%%

We first briefly review the basic concept of Conformal Fermi Coordinates (CFC), which was first introduced in \cite{Pajer:2013ana} and defined
rigorously in \cite{Dai:2015rda}.  Consider an observer
free-falling in some spacetime.  We are mostly considering applications
in the cosmological context, in which this spacetime is approximately
FLRW, but this does not have to be the case.  Her trajectory
is a timelike geodesic $G$, whose tangent vector we will denote as $e_0$.  One
can construct a frame where the spatial origin is always located on this
geodesic, and in which the metric takes the form\footnote{Here we have let $a_F$ be a function of the CFC time coordinate $\tau_F$ only.  While this is not strictly necessary (see \cite{Dai:2015rda}), this is the most natural choice as constant-scale-factor surfaces then coincide with constant-proper-time surfaces.}  
\bea
\label{eq:CFNC-requirement}
g^F_{\mu\nu}(x^\mu_F) = a^2_F(\tau_F) \left[ - \eta_{\mu\nu} + h^F_{\mu\nu} (\tau_F, x^i_F) \right],\qquad h^F_{\mu\nu} = \mathcal{O}[(x^i_F)^2]\, .
\eea 
Thus, in these coordinates the metric looks like an unperturbed FLRW metric
in the vicinity of the observer's trajectory \emph{at all times}, 
up to corrections that go as the spatial distance from the origin squared.  
Note that a metric of the form \refeq{CFNC-requirement} means that an observer at
the origin $x_F^i = 0$ is free-falling, with proper time $t_F$ given by
\be
t_F(\tau_F) = \int_0^{\tau_F} a_F(\tau') d\tau'\,.
\ee
Briefly, this frame can be constructed as follows.  For now, let $a_F(x)$
be a positive scalar field in the neighborhood of G.  We introduce the conformal metric
\be
\tilde g_{\mu\nu}(x) = a_F^{-2}(x) g_{\mu\nu}(x)\,.
\ee
We then define a ``conformal proper time'' $\tau_F$ along $G$ through
\bea
dt_F = a_F(P) d\tau_F\,,
\eea
where $P$ denotes the point on the central geodesic which has proper time
$t_F$.  This defines our time coordinate $x_F^0 = \tau_F$.  At some point
$P$ along $G$ we can construct an orthonormal tetrad $(e_a)$, where $e_0 = \partial/\partial t_F$
is again the tangent vector to $G$, with $(e_a)^\mu (e_b)^\nu g_{\mu\nu} = \eta_{ab}$.  
The tetrad is defined at all other points on $G$ by parallel transport, so that
this condition is preserved.  A point $Q$ corresponding to CFC coordinates
$x_F^\mu$ is then located as follows.  First, we move to the point $P$ on $G$ specified by $x_F^0$.  We 
then construct a spatial geodesic $\tilde h(x_F^0; x_F^i; \lambda)$ 
of the \emph{conformal} metric 
$\tilde g_{\mu\nu}$ which satisfies $\tilde h(x_F^0; x_F^i; 0) = P$, and  
whose tangent vector at $P$ is given by
\be
\frac{\partial}{\partial\lambda} \tilde h^\mu(x_F^0; x_F^i; \lambda) \Big|_{\lambda=0} = \frac{x_F^i}{s} (e_i)^\mu;\quad  s = (\delta_{ij} x_F^i x_F^j)^{1/2}\,.
\ee
The point $Q$ is then found by following this geodesic for a \textit{proper distance} $\lambda = a_F(P) s$. As we show in \cite{Dai:2015rda}, this yields a metric in the form \refeq{CFNC-requirement}.  Moreover, the leading corrections are given by
\ba
h^F_{00} =\:& - \tilde R^F_{0l0m} x_F^l x_F^m \vs
h^F_{0i} =\:& -\frac23 \tilde R^F_{0lim} x_F^l x_F^m \vs
h^F_{ij} =\:& -\frac13 \tilde R^F_{iljm} x_F^l x_F^m\,,
\label{eq:hF_general}
\ea
where $\tilde R^F_{\alpha\beta\gamma\delta}$ is the Riemann tensor of the conformal
metric, evaluated in the CFC frame at $x_F^0$ on the central geodesic. 

Note that if we now were to choose $ a_{F}=1 $, the CFC would reduce exactly to the ordinary Fermi Normal Coordinates \cite{Manasse:1963zz}.  Then, the metric perturbations $h^F_{\mu\nu}$ would contain the leading corrections due to Hubble flow $\propto H^2 r_F^2$, restricting the validity of the coordinates to subhorizon scales.  Instead, we will choose $ a_{F} $ so that it captures the locally observable expansion of the Universe, which automatically extends the validity of the coordinates to the spatial scale of perturbations which can be superhorizon.  The well-defined prescription that fixes $ a_{F} $ up to a multiplicative constant will be given in \refsec{perturbed-FLRW}.  

\comment{Alternatively, the CFC metric can be derived via the usual tensorial transformation rule. This can be re-cast into the following form
\bea
\label{eq:CFNC-metric-conformal-transform}
\left[ (a_F(P))^{-2} g_{F,\mu\nu}\right] (x^\lambda_F) = \left[ \frac{a(P)}{a_F(P)} \frac{\partial x^\alpha}{\partial x^\mu_F} \right] \left[ \frac{a(P)}{a_F(P)} \frac{\partial x^\beta}{\partial x^\nu_F} \right] \left( \frac{a(\tau)}{a(P)} \right)^2 \left[ \left(a(\tau)\right)^{-2} g_{\alpha\beta} \right] (x^\lambda),
\eea
where we have massaged with scale factors in both coordinate systems. This can be instead understood as how the conformal metric $a^{-2}g_{\mu\nu}$ transforms. A subtle point here, is that different points on a constant-$\tau_F$ surface (and hence correspond to the same $a_F(\tau_F)$ in CFC are not simultaneous in the global coordinates. Therefore, $a(\tau)$ and $a(P)$ are numerically different.}

%%%%%%%%%%%%%%%%%%%%%%%%%%%%%%%%%%%%%%%%%%%%%%%%%%%%%%%%%%%%%%%%
\subsection{Scalar perturbations around an FLRW universe}

%%%%%%%%%%%%%%%%%%%%%%%%%%%%%%%%%%%%%%%%%%%%%%%%%%%%%%%%%%%%%%%% 

We now turn to an FLRW universe with scalar perturbations in the conformal-Newtonian (cN) gauge, 
\bea
\label{eq:perturbed-FLRW}
ds^2 = a^2(\tau) \left[ - (1+2\Phi) d\tau^2 + (1-2\Psi) \delta_{ij} dx^i dx^j\right]\,.
\eea
In the absence of anisotropic stress we have $\Psi=\Phi$; however, we will not make that assumption in this section.  Before we apply the CFC construction, we note that
we are implicitly performing a coarse-graining of the metric \refeq{perturbed-FLRW} on some scale $\Lambda$.  This is because in the actual universe, scalar
perturbations exist on all scales, so that without any coarse-graining the metric \refeq{CFNC-requirement}, which assumes that the corrections $h^F_{\mu\nu}$ are small, is only valid on infinitesimally small scales.  In the following, $\Phi$ and $\Psi$ will thus denote coarse-grained metric perturbations.  

Further, we will work to linear order in $\Phi$ and $\Psi$.  
The perturbative expansion in $\Phi,\,\Psi$ should not be confused with the power expansion in $x^i_F$. The former expansion is valid as long as $|\Phi,\,\Psi| \ll 1$, and is chosen here for simplicity; the CFC construction also works for spacetimes that differ strongly from FLRW.  The latter expansion on the other hand is good if $|x^i_F|$ is smaller than the typical variation scale for $\Phi,\,\Psi$, which is the fundamental expansion parameter in CFC.  Following 
standard convention, spacetime indices $\mu,\nu,...$ are assumed to be
raised and lowered with the metric $g_{\mu\nu}$, while latin indices
$i,j,k,...$ are raised and lowered with $\delta_{ij}$.  Correspondingly,
$\partial^2$ denotes the flat-space Laplacian
\be
\partial^2 = \delta^{ij} \partial_i \partial_j\,.
\ee

%%%%%%%%%%%%%%%%%%%%%%%%%%%%%%%%%%%%%%%%%%%%%%%%%%%%%%%%%%%%%%%% 
\subsubsection{CFC construction}
\label{sec:perturbed-FLRW}
%%%%%%%%%%%%%%%%%%%%%%%%%%%%%%%%%%%%%%%%%%%%%%%%%%%%%%%%%%%%%%%% 

We begin with deriving the tetrad.  Consider a free-falling observer traveling along the time-like central geodesic $G$. His 4-velocity is given by 
\be
U^\mu = (e_0)^\mu = a^{-1} \left(1- \Phi, V^i \right)\,, 
\ee
where the 3-velocity $V^i$ is considered as first-order perturbation.\footnote{Throughout, $V^i$ will denote the large-scale peculiar velocity in global coordinates, while we will later encounter the small-scale peculiar velocity $v^i$.}
$V^i$ obeys the geodesic equation, given by
\bea
\label{eq:CFNC-obs-eom}
V^{i\prime} + \coH V^i & = & - \partial^i \Phi \,,
\eea  
where a prime denotes derivative with respect to conformal time $\tau$.  
Our choice of the spatial components of the orthonormal tetrad is
\bea
(e_j)^\mu & = & a^{-1} \left( V_{j}, \delta^i{}_j [1+\Psi] \right)\,,
\eea 
which we have aligned with the global coordinate axes for simplicity and without loss of generality.  

We now want to determine $a_F$ by defining the locally observable expansion rate of the Universe to match the CFC Hubble rate $\coH_F \equiv a_F^{-1} da_F/d\tau_F$.  Following \cite{Dai:2015rda}, consider the {\it convergence of the geodesic congruence} $\nabla_\mu U^\mu$ along the central geodesic $G$. This is a locally measurable quantity, which intuitively corresponds to the change in time of the proper volume of a bundle of geodesic trajectories in the neighborhood of the central geodesic $G$.  In
an unperturbed FLRW universe,  we have
\be
\nabla_\mu U^\mu = 3 H = 3 \coH / a,.
\ee
Thus, we define at all points along the geodesic $G$
\be
H_F(\tau_F) \equiv \frac{\coH(\tau_F)}{a_F(\tau_F)} = \frac13 \nabla_\mu U^\mu\Big|_G\,.
\label{eq:congruenceF}
\ee
In terms of cN gauge quantities the convergence of a geodesic congruence is given by
\bea
\label{eq:congruence-central-geodesic}
\nabla_\mu U^\mu = 3 H \left[ 1 - \Phi - \coH^{-1} \Psi' + (3\coH)^{-1} \partial_i V^i \right] \,.
\eea
The local Hubble parameter %$H_F = \coH_F/a_F$ 
is then given by
\bea
\label{eq:local-Hubble}
H_F = H \left[ 1 - \left( \Phi + \frac{1}{\coH} \Psi' \right) + \frac{1}{3\coH} \partial_i V^i \right].
\label{eq:HF}
\eea
Note that Ref.~\cite{Baldauf:2011bh} in their construction using the standard Fermi Normal Coordinates (FNC) only include the last, velocity-divergence term, which dominates when the long-wavelength mode is deep inside the horizon.  However, the second term here is a necessary correction on (super-)horizon scales, since on those scales $V^i$ does not correspond to physical peculiar velocities.   

Consider, for example, the motion of two nearby fluid elements in the presence of a superhorizon long-wavelength mode. For adiabatic initial conditions, the two fluid elements will then have vanishing relative peculiar velocity, which scales as $(k/\coH)^2 \Phi$, i.e.~the fluid elements have a constant, infinitesimal coordinate separation $\Delta x$. The physical separation is $d_{\rm phys} = a (1-\Psi) \,\Delta x$.  The proper time interval is $a(1+\Phi)d\tau$. The physical relative velocity $v_{\rm phys}$, i.e.~the rate of change of the proper distance between them, then reads
\bea
v_{\rm phys} = \frac{1}{a(1+\Phi)} \frac{d }{d\tau} d_{\rm phys} & = & \frac{1}{a(1+\Phi)} \frac{d}{d\tau} \left[ a (1-\Psi)\, \Delta x \right] = \coH \left( 1 - \Phi - \Psi - \frac{1}{\coH} \Psi' \right)\, \Delta x \nn\\
& = & H \left[ 1 - \left( \Phi + \frac{1}{\coH} \Psi' \right) \right]\, a (1-\Psi)\, \Delta x = H_F \, d_{\rm phys}\,,
\eea
where the last equality holds since we neglect terms of order $(k/\coH)^2\Phi$.  
Therefore, the relative velocity $v_{\rm phys}$ and $H_F$ satisfy exactly the local version of Hubble's law, showing that \refeq{HF} is indeed the proper expression for the local Hubble rate.  

Since so far we have only fixed $H_F = d\ln a_F/dt_F$, the local scale factor $a_F$ is defined up to a multiplicative constant.  In other words, we have a residual freedom to rescale $\{ a_{F},\tau_{F}\} \rightarrow \{a_{F}\lambda,\tau_{F}/\lambda\} $, which keeps the proper time $ t_{F} $ unchanged.
Of course, this constant is arbitrary and should cancel out of any proper
observable.  We will here fix the constant in order to make our results
of the next sections more transparent.  Specifically, we demand that
\be
\frac{a_F(\tau_F)}{a(\tau_F)} \rightarrow 1 \qquad {\rm as} \quad \tau_F \rightarrow 0\,,
\ee
which means that at early times, the local scale factor-proper time relation is the same as that in the \emph{unperturbed} background cosmology.  
The ratio of scale factors at a fixed \emph{spacetime} point is then
at early times given by
\bea
\lim_{\tau_F\to 0} \frac{a_F(\tau_F)}{a(\tau(\tau_F))} = 1 + \frac23 \Phi_{\rm ini}\,,
\label{eq:aFIC}
\eea
where $\Phi_{\rm ini}$ is the asymptote as $\tau_F\to 0$ of the potential along the geodesic.  
A direct integration gives
\bea
\frac{a_F(\tau_F)}{a(\tau)} & = & 1 + \frac23 \Phi_{\rm ini} + \int^\tau_0 d\tau' \left( - \Psi' + \frac13 \partial_i V^i \right)\,.
\label{eq:aFovera}
\eea
In order to obtain the leading corrections to the CFC metric (either via
\refeq{hF_general} or an explicit coordinate transformation of the metric),
we also need the second derivative $a''_F$.  This can be obtained easily
from \refeq{congruenceF},
\bea
\label{eq:eq-CFNC-acceleration}
\frac{1}{a^2_F(\tau_F)} \frac{d\coH_F(\tau_F)}{d\tau_F} 
= \frac13 \frac{\coH_F(\tau_F)}{a_F(\tau_F)} \nabla_\mu U^\mu + \frac13 (e_0)^\nu \,\nabla_\nu \nabla_\mu U^\mu\,,
\eea
where again all quantities are evaluated on $G$.

%%%%%%%%%%%%%%%%%%%%%%%%%%%%%%%%%%%%%%%%%%%%%%%%%%%%%%%%%%%%%%%%
\subsubsection{CFC metric}
\label{sec:long-scalar}
%%%%%%%%%%%%%%%%%%%%%%%%%%%%%%%%%%%%%%%%%%%%%%%%%%%%%%%%%%%%%%%% 

We can now derive the remaining corrections $h^F_{\mu\nu}$ to the CFC 
metric, making use of the geodesic equation for $V^i$.   This yields in terms
of the conformal Newtonian gauge perturbations
\bea
\label{eq:eq-187}
h^F_{00} & = & - \left( \partial_k \partial_l \Phi - \frac13 \delta_{kl} \partial^2 \Phi \right)\, x_F^k x_F^l \\
\label{eq:eq-188}
h^F_{0i} & = & \frac23 \left( \coH' - \coH^2 \right) \left( \delta_{kl}  V_i - \delta_{il} V_k \right)\, x_F^k x_F^l + \frac23 \left( \delta_{ki} \partial_l - \delta_{kl} \partial_i \right) \left( \Psi' + \coH \Phi \right)\, x_F^k x_F^l \\
\label{eq:eq-189}
h^F_{ij} & = & \frac13 \left( \delta_{lj} \partial_i \partial_k \Psi + \delta_{li} \partial_j \partial_k \Psi - \delta_{ij} \partial_k \partial_l \Psi - \delta_{kl} \partial_i \partial_j \Psi \right)\, x_F^k x_F^l \\
& & + \frac29 \coH\, \partial^2 V \left( \delta_{ij} \delta_{kl} - \delta_{ik} \delta_{jl} \right)\, x_F^k x_F^l\,.\nonumber
\eea
Let us consider a spherically symmetric configuration about the CFC observer. It amounts to setting $V^i$, $\partial_i \Phi$ and $\partial_i \Psi$ to zero, and 
replacing $\partial_i \partial_j \Phi$ with $(1/3) \delta_{ij} \partial^2 \Phi$ and the same for $\Psi$.  We then find that $h^F_{00}$ and $h^F_{0i}$ vanish,
while $h^F_{ij}$ reduces to
\bea
\label{eq:eq-3.13-KF}
h^F_{ij} = \frac13 K_F \left( x_{F,i} x_{F,j} - \delta_{ij}\, r_F^2 \right)\,,
\eea 
where we have suggestively introduced
\be
K_F \equiv \frac23 \left(\partial^2 \Psi - \coH \partial_i V^i \right)\,.
\label{eq:KFdef}
\ee 
The tensorial structure $\propto \left(x_F^i x_F^j - \delta^{ij} r_F^2 \right)$ distorts the proper distance only along directions perpendicular to the radial direction, as guaranteed by our construction of CFC (recall that $r_F$ is defined as the geodesic distance in the conformal metric).  We are free to choose a different radial coordinate (see \refapp{spatialgauge} for a discussion of the residual gauge freedom in the spatial component of the CFC metric)
\bea
\label{eq:eq-106}
r_F \longrightarrow r_F \left( 1 - \frac{1}{12} K_F\, r_F^2 \right), \qquad dr_F \longrightarrow \left( 1 - \frac14 K_F\, r_F^2 \right) dr_F\,.
\eea
In terms of this new radial coordinate, the area radius of the spacetime (see \refsec{proof}) is given by $(1+K_F\,r^2_F/4)^{-1} r_F$. Under this rescaling, the CFC metric simply becomes (valid to $\mathcal{O}[(x^i_F)^2]$)
\bea
\label{eq:stereographic-curved-FLRW}
ds^2 = a_F^2(\tau_F) \left[ - d\tau_F^2 + \frac{ \delta_{ij} dx_F^i dx_F^j}{\left(1+K_F\, r_F^2/4\right)^2} \right].
\eea
The spatial part of the metric is the familiar stereographic parameterization of a curved, homogeneous space.

We now go back to general anisotropic configurations of $\Phi,\,\Psi$. 
In the next section, we will see that, in the context of scalar perturbations considered here, it is necessary that all cosmic fluids are comoving with velocity $V_i$ for the separate universe picture to hold.  The comoving condition implies that $h^F_{0i}$ in fact vanishes, by way of the time-space Einstein equation. Therefore, we will set $h^F_{0i}$ to zero in the remainder of the paper.
We can again use the residual gauge freedom described in \refapp{cN-CFNC}
to bring $h^F_{ij}$ into a more familiar form.  After some algebra, this yields
\ba
ds^2 = a_F^2(\tau_F) \Big\{ & - \left[ 1 + \left( \partial_k \partial_l \Phi - \frac13 \delta_{kl} \partial^2 \Phi \right) x_F^k x_F^l \right] d\tau_F^2 \vs
&
+ \left[ 1 - \left( \partial_k \partial_l \Psi - \frac13 \delta_{kl} \partial^2 \Psi \right) x_F^k x_F^l \right] \frac{\delta_{ij} dx_F^i dx_F^j}{\left(1+K_F\, r^2/4\right)^2} \Big\}\,.
\label{eq:hFgeneral}
\ea
We have thus put the CFC metric into the conformal Newtonian form.  
The metric perturbations $\Phi,\,\Psi$ enter in two distinct ways.  First,
the isotropic part of the perturbation (proportional to $\partial^2\Psi$ and
$\Phi,\,\Psi$ themselves as well as their time derivatives) modifies the background $a_F(\tau_F)$ and leads
to spatial curvature.  The anisotropic part, which is completely determined
by
\be
K^{\Phi}_{ij} \equiv \left( \partial_i \partial_j  - \frac13 \delta_{ij} \partial^2 \right) \Phi\,,
\ee
and the analogous $K^{\Psi}_{ij}$, enters as a tidal field.  Note that
since the CFC frame was constructed based on local observables, no
gauge modes yield any observable imprint in the CFC metric.  
Further, it is important to emphasize that no subhorizon assumption
has been made about either $\Phi,\,\Psi$ or the CFC metric itself; 
\refeq{hFgeneral} is valid on arbitrarily large scales as long as the
corrections are small.  
We will turn to the interpretation of the CFC metric in the next section.

%%%%%%%%%%%%%%%%%%%%%%%%%%%%%%%%%%%%%%%%%%%%%%%%%%%%%%%%%%%%%%%%%%%%%%%%%%%%%%
%%%%%%%%%%%%%%%%%%%%%%%%%%%%%%%%%%%%%%%%%%%%%%%%%%%%%%%%%%%%%%%%%%%%%%%%%%%%%%
\section{Separate universe revisited}
\label{sec:local-friedmann-eqs}
%%%%%%%%%%%%%%%%%%%%%%%%%%%%%%%%%%%%%%%%%%%%%%%%%%%%%%%%%%%%%%%%%%%%%%%%%%%%%%
%%%%%%%%%%%%%%%%%%%%%%%%%%%%%%%%%%%%%%%%%%%%%%%%%%%%%%%%%%%%%%%%%%%%%%%%%%%%%%

In the previous section we derived the CFC frame metric for scalar perturbations, which \emph{appears} to be of the FLRW form with tidal corrections.  
However, in order to test whether an observer along the central geodesic 
would indeed interpret local measurements as an FLRW universe in the absence of tidal corrections, we need
to derive the equation for $a_F$ and compare it to the Friedmann equations.  
Further, we need to verify that the ``curvature'' $K_F$ is indeed 
constant.  Throughout we will set the tidal corrections in \refeq{hFgeneral} to zero in keeping with our linear order treatment.

Let us consider an observer traveling along the central geodesic who performs
local measurements.  Suppose further this observer knows about General Relativity.  
She will compare the local Hubble rate $H_F$ with the Friedmann equations, the first of which reads
\bea
H^2_F = \frac{8\pi G}{3}  \rho_F - \frac{K^{\rm Fr}_F}{a^2_F} \,,
\label{eq:Fr1}
\eea
where $\rho_F$ is the local rest-frame matter density, coarse-grained over 
the region considered as CFC patch.  Using \refeq{local-Hubble}, 
$K_F^{\rm Fr}$ is given in terms of global quantities by
\bea
\label{eq:eq-KF}
K_F^{\rm Fr} = \coH^2 \Delta_{\rm cN} + 2 \coH \left( \Psi' + \coH \Phi \right) -  \frac23 \coH\, \partial_i V^i\,,
\eea
where $\Delta_{\rm cN}$ is the density perturbation in cN gauge.  
On the other hand, the spatial curvature $K_F$ 
which can be inferred from, e.g. shining light rays and
measuring angles between them, is given in terms of global quantities by \refeq{KFdef}.  We thus need to verify whether $K_F^{\rm Fr} = K_F =$~const, and under which conditions.  Only in that case would the observer actually interpret
the results of local measurements as an FLRW universe.  

Let us assume that the Universe is filled with multiple uncoupled fluids labeled by $I=1,2,\cdots$. Each fluid has a homogeneous equation of state $\bar{\mathcal{P}}_I = w_I\bar\rho_I$, with possibly time-dependent $w_I$. The homogenous part of each fluid evolves as $\bar\rho'_I=-3\coH(1+w_I) \bar\rho_I$, which implies
\bea
\label{eq:eq-C1}
\left(\coH^2 \Omega_I\right)' = -\left(1+3w_I\right) \coH^3 \Omega_I,
\eea
where $\Omega_I=\bar\rho_I/\bar\rho$ and $\bar\rho$ is the total (homogeneous) energy density. The acceleration is given by
\bea
\label{eq:eq-C2}
\coH' = - \frac12 \coH^2 \left( \sum_I (1+3w_I) \Omega_I \right).
\eea
In the global cN-gauge coordinates, we let each component have a fractional density perturbation $\Delta_I$ and
peculiar velocity $V_I^i = \partial^i V_I$, where $V_I$ is the velocity potential.  
Note that $V^i$ denotes the velocity of the central geodesic throughout.  
We now write down the time-time and time-space Einstein equations, which have the sum of all fluids as sources,
\bea
\label{eq:eq-C3}
\partial^2 \Psi & = & \frac32 \coH^2 \left( \sum_I \Omega_I \Delta_I \right) - \frac92 \coH^3 \left( \sum_I \Omega_I \left(1+w_I\right) V_I \right), \\
\label{eq:eq-C4}
\Psi' + \coH \Phi & = & - \frac32 \coH^2 \left( \sum_I \Omega_I \left(1+w_I\right) V_I \right).
\eea
Using that
\be
%\frac{\rho_F-\rhob}{\rhob} 
\Delta_{\rm cN} = \sum_I \Omega_I \Delta_I\,,
\ee
\refeqs{eq-C3}{eq-C4} can be inserted into the first two terms of \refeq{eq-KF} to yield
\be
K_F^{\rm Fr} = \frac23 \left(\partial^2 \Psi - \coH \partial_i V^i \right) = K_F\,.
\ee
This is \refeq{KFdef}, showing that the quantity appearing in the local
Friedmann equation is indeed the spatial curvature.  Thus, \refeq{eq-KF} provides an equivalent relation between the long-wavelength adiabatic perturbation in 
conformal-Newtonian gauge and the local curvature.

%%%%%%%%%%%%%%%%%%%%%%%%%%%%%%%%%%%%%%%%%%%%%%%%%%%%%%%%%%%%%%%%%%%%%%%%%%%%%%%
\subsection{Conservation of curvature}
\label{sec:KF}
%%%%%%%%%%%%%%%%%%%%%%%%%%%%%%%%%%%%%%%%%%%%%%%%%%%%%%%%%%%%%%%%%%%%%%%%%%%%%%%

In the FLRW solution, the spatial curvature $K_F$ is constant.  This is thus a necessary condition that needs to be satisfied for the separate universe picture to hold.  To study this, we allow each fluid to have a pressure perturbation $\delta\mathcal{P}_I$ in addition to the density perturbation $\delta\rho_I=\rho_I \Delta_I$. In the rest frame of the fluid, we have
\bea
\delta\mathcal{P}_I \equiv c^2_{s,I}\, \delta\rho_I,
\eea
where the rest-frame speed of sound $c^2_{s,I}$ does not necessarily equal the adiabatic sound speed
\bea
c^2_{{\rm ad},I} \equiv \frac{\bar{\mathcal{P}}'_I}{\bar\rho'_I} = w_I - \frac{w'_I}{3\coH (1+w_I)},
\eea 
because the pressure perturbation can be non-adiabatic. Each fluid then evolves according to~\cite{Bean:2003fb}
\bea
\label{eq:eq-C5}
\Delta'_I + \left(1+w_I\right) \partial^2 V_I - 3 \left(1+w_I\right) \Psi' & = & 3 \coH \left( w_I - c^2_{s,I} \right) \left[ \Delta_I - 3 \coH (1+w_I) V_I \right] \vs
& & + 3 \coH w'_I V_I \\
%%%
\label{eq:eq-C6}
V'_I + \left(1 - 3 c^2_{s,I} \right) \coH V_I + \frac{c^2_{s,I}}{1+w_I} \Delta_I + \Phi & = & 0\,.
\eea
Using \refeqs{eq-C1}{eq-C2}, \refeqs{eq-C3}{eq-C4}, and \refeqs{eq-C5}{eq-C6} (keeping $c^2_{s,I}$ general), we can directly compute the time derivative of \refeq{eq-KF}.\footnote{Note that $d/d\tau=\partial_\tau + V^i \partial_i$ is in principle the total derivative along the central geodesic; however, the spatial derivative term has been neglected as it is higher order in perturbations.}  We also assume that $V^i$ follows the peculiar motion of the non-relativistic matter and therefore satisfies the geodesic equation (the $c^2_{s,I}=0$ version of \refeq{eq-C6}). We obtain
\bea
\label{eq:dKFdtau}
\frac{d K_F}{d\tau}
& = & \left[ \coH^2 \left( \sum_I \Omega_I  \Delta_I \right) - 3 \coH^3 \left( \sum_I \Omega_I (1+w_I) V_I \right) -  \frac23 \coH \,\partial_i V^i \right]' \vs
& = & - \coH^2 \left[ \sum_I \left( 1 + w_I \right) \Omega_I \partial_i \left( V^i_I - V^i \right) \right] + \frac23 \coH\, \partial^2 \left( \Phi - \Psi \right).
\eea
The final result is obtained after lengthy but straightforward algebra. The second term vanishes if long-wavelength perturbations do not source anisotropic stress $\Phi=\Psi$.

If the Universe is only composed of non-relativistic matter ($w=c^2_s=0$) plus a possible cosmological constant ($w=-1$), and the CFC observer co-moves with the matter fluid $V=V_m$ along the central geodesic, $K_F$ is conserved and the observer cannot distinguish the spacetime from a curved $\Lambda$CDM universe using any local measurements (that is, measurements over a scale much smaller than the wavelength of the perturbations $\Phi,\,\Psi$).  This applies to any isotropic configuration around the geodesic, and is not restricted to the tophat considered in \refsec{proof}.

If there are extra fluids with other values of $w_I$ in the Universe, as postulated in many of the alternative cosmologies, $dK_F/d\tau$ in general does not evaluate to zero. If all fluids co-move with the CFC observer $V_I^i = V^i$, however, $K_F$ does remain a constant.  Since this situation requires all fluids to follow geodesic motion, non-gravitational forces have to be negligible. This will be true when the sound horizon is much smaller than the wavelength of interest, i.e.~$\int a(\tau) c_s(\tau) d\tau \ll a/k_L$~\cite{Creminelli:2009mu}.  This further generalizes our considerations of \refsec{proof}.

Assuming that the CFC observer co-moves with the average cosmic fluid such that $V_{i} = \delta T^0{}_i / (\bar\rho + \bar{\mathcal{P}})$, we can re-write \refeq{eq-KF} using \refeqs{eq-C3}{eq-C4} to find
\bea
\label{eq:KF-comoving-curvature}
K_F = \frac23 \left( \partial^2 \Psi - \frac{\coH\,\partial^i \delta T^0{}_i}{\bar\rho+\bar{\mathcal{P}}} \right) = \frac23 \partial^2 \mathcal{R},
\eea
where $\mathcal{R}$ is the gauge-invariant {\it curvature perturbation on comoving slices}. Notice that we have made no assumption about the decaying adiabatic mode (the second adiabatic solution, always present in the small $ k $ limit \cite{Weinberg:2003sw}). Its contribution cancels out of this equation. The conservation of $K_F$ is hence directly related to the conservation of $\mathcal{R}$ and it is valid also in the presence of a decaying adiabatic mode. Note that the conditions for $\mathcal{R} =$~const are different than for $\zeta=$~const, where $\zeta$ is the curvature perturbation on \emph{uniform density slices}.  The latter is conserved if pressure is only a function of energy density~\cite{Lyth:2004gb}, while the conservation of $K_F$ does not exclude non-adiabatic pressure.  Thus, $\mathcal{R}$, not $\zeta$, is the locally observable curvature, and the separate universe picture relies on conservation of the former, not the latter.   
Note that in \cite{Baldauf:2011bh,Wagner:2014aka}, the comoving
curvature perturbation was denoted with $\zeta$, specifically $\mathcal{R}_{\rm here} = -\zeta_{\rm there}$.  Thus, our result \refeq{KF-comoving-curvature} agrees with those references.  \refeq{KF-comoving-curvature}, together with the discussion after \refeq{HF} above, thus confirms that superhorizon perturbations cannot affect the cosmology as observationally determined by measurements within the horizon~\cite{hirata/seljak:05}.  

Finally, one can also verify that the second Friedmann equation is 
satisfied by $H_F$, which is shown in \refapp{Fr2}.  
Unlike the first Friedmann equation, this does not
impose any conditions on the fluid components. Indeed, it is a simple
consequence of the Raychaudhuri equation applied to the CFC metric \refeq{hFgeneral}.

%%%%%%%%%%%%%%%%%%%%%%%%%%%%%%%%%%%%%%%%%%%%%%%%%%%%%%%%%%%%%%%%%%%%%%%%%%%%%%%
\subsection{Relation to synchronous-comoving gauge}
\label{sec:sc}
%%%%%%%%%%%%%%%%%%%%%%%%%%%%%%%%%%%%%%%%%%%%%%%%%%%%%%%%%%%%%%%%%%%%%%%%%%%%%%%

It is illuminating to connect the curvature $K_F$, \refeq{KF-comoving-curvature}
to the synchronous-comoving (sc) gauge.  Following \cite{Bertschinger:1993xt},
we write the metric in global coordinates as
\be
ds^2 = a^2(\tau) \left[-d\tau^2 + \left(\frac13 h \delta_{ij} + \mathcal{D}_{ij} (\partial^{-2} \xi)\right) dx^i dx^j \right]\,,
\label{eq:scgauge}
\ee
where $\mathcal{D}_{ij} = \partial_i\partial_j - \delta_{ij}\partial^2/3$.  
Here and in the following, we drop the subscript $F$ on the spatial CFC coordinates,
since the global coordinates will not appear explicitly anymore and so there
is no ambiguity.  
The time-time component of the Einstein equation is
\be
- \frac16 \partial^2 (h-\xi) + \frac12 \coH h' = 4\pi G \bar\rho a^2 \Delta_{\rm sc}\,,
\label{eq:G00sc}
\ee
where $\Delta_{\rm sc}$ is the density perturbation in synchronous-comoving gauge.  Further, the comoving condition $V^i=0$ implies that $\xi'=h'$.  Note
that the quantity on the l.h.s. of \refeq{G00sc} is $\partial^2 \Psi$, i.e.
we have
\be
\partial^2\Psi = 4\pi G\bar\rho a^2 \Delta_{\rm sc}\,.
\label{eq:Poissonsc}
\ee
That is, the spatial metric potential in cN gauge is
related to the density perturbation in sc gauge by
the Newtonian Possion equation.  This relation holds on all scales, a fact
which is being used in N-body simulations \cite{Chisari:2011iq}.  

One can easily verify that the local Hubble rate $H_F$ for the metric
\refeq{scgauge} is given by
\be
H_F = \frac13 \nabla_\mu U^\mu = H + \frac12 a^{-1} h'\,.
\label{eq:HFsc}
\ee
Let us now consider a spherically symmetric perturbation in sc gauge. Straightforward evaluation of the
spatial part of the conformal Riemann tensor, together with $\partial_i\partial_j h = \delta_{ij} \partial^2h/3$ then yields
\be
h_{ij}^F = - \frac{1}{27} \partial^2 (h-\xi)  \left( x_i x_j - \delta_{ij} r^2 \right)\,,
\ee
and thus, matching to the spatial curvature (\refeq{eq-3.13-KF}),
\be
K_F = -\frac19 \partial^2 (h-\xi)\,.
\ee
Let us verify again the local Friedmann equation,
\be
H_F^2 = \frac{8\pi G}{3} \rho_F - \frac{K_F}{a_F^2}
\ee
Multiplying the Friedmann equation by $a^2$, using \refeq{HFsc}, and subtracting the background contribution, we obtain
\be
\frac13 \coH h' = \frac{8\pi G}{3} a^2 \rhob \Delta_{\rm sc} + \frac19 \partial^2 (h-\xi)\,,
\ee
which is exactly the time-time component of the Einstein equation \refeq{G00sc} (multiplied by $2/3$).  

Let us now restrict to an EdS universe.  The continuity equation
for pressureless matter in sc gauge reads~\cite{Bertschinger:1993xt}
\be
\Delta_{\rm sc}' = - \frac12 h'\,.
\ee
We thus have
\ba
K_F =\:& \frac{8\pi G}{3} a^2 \rhob \Delta_{\rm sc} - \frac13 \coH h'
= \coH^2 \Delta_{\rm sc} + \frac23 \coH \Delta_{\rm sc}' \vs
=\:& \frac53 H_0^2 a^{-1} \Delta_{\rm sc}\,,
\ea
where we have used that $\Delta_{\rm sc} \propto a(\tau)$ while $\coH^2 = H_0^2 a^{-1}$ for EdS universe.  
Note that for any (spatially flat) cosmology where $K_F$ is a constant and which has an epoch
of matter domination, during which $H^2(\tau) = (\Omega_{m0}/\Omega_m) H_0^2 a^{-3} \stackrel{\rm MD}{\to} \Omega_{m0} H^2_0 a^{-3}$,
this expression can be simply evaluated during that epoch, so that
for any later expansion history ($\Lambda$CDM or other)
\be
\frac{K_F}{H_0^2} = \frac53 \Omega_{m0} \frac{\Delta_{\rm sc}(\tau)}{D(\tau)}\,,
\label{eq:KFsc}
\ee
where $D(\tau)$ is the linear growth factor normalized to $a(\tau)$ during
matter domination.  Thus, the local curvature is directly related (during matter domination)  to
the matter density perturbation \emph{in synchronous-comoving gauge}.  
Moreover, we find the same relation as in the nonlinear compensated
tophat of \refsec{proof}, \refeq{Ki}.

%%%%%%%%%%%%%%%%%%%%%%%%%%%%%%%%%%%%%%%%%%%%%%%%%%%%%%%%%%%%%%%%%%%%%%%%%%%%%%%
%%%%%%%%%%%%%%%%%%%%%%%%%%%%%%%%%%%%%%%%%%%%%%%%%%%%%%%%%%%%%%%%%%%%%%%%%%%%%%%
\section{Application: the squeezed-limit matter bispectrum on large scales}
\label{sec:app-squeezed-matter-bispec}
%%%%%%%%%%%%%%%%%%%%%%%%%%%%%%%%%%%%%%%%%%%%%%%%%%%%%%%%%%%%%%%%%%%%%%%%%%%%%%%
%%%%%%%%%%%%%%%%%%%%%%%%%%%%%%%%%%%%%%%%%%%%%%%%%%%%%%%%%%%%%%%%%%%%%%%%%%%%%%%

Let us now turn to an application of the CFC frame for long-wavelength
scalar perturbations, namely calculating the physical influence of 
these perturbations on the growth of small-scale structure. 
We will restrict to non-relativistic pressure-free matter on short scales. 
As described
in detail in \cite{Dai:2015rda}, we coarse-grain the metric on some spatial
scale $L = \Lambda^{-1}$, and perform the CFC construction with respect
to this coarse-grained metric.  In \cite{Dai:2015rda}, we used the notation
$h^\Lambda$ for the long-wavelength part of the metric perturbation,
which contains only Fourier modes with $k_L \lesssim \Lambda$, while the
remaining short-wavelength metric was denoted $h^s$, with support for
$k_s \gtrsim \Lambda$ in Fourier space.  Here, for clarity of notation and
since we have adopted the conformal-Newtonian gauge, we will let $\Phi,\,\Psi$
stand for the long-wavelength perturbation, while $\phi,\,\psi$ denote
the small-scale metric potentials.  
In the CFC frame, the metric
is then a simple extension of \refeq{hFgeneral},
\ba
ds^2 = a_F^2(\tau_F) \Bigg\{ & - \left[ 1 + 2\phi + \left( \partial_k \partial_l \Phi - \frac13 \delta_{kl} \partial^2 \Phi \right) x^k x^l \right] d\tau_F^2 \vs
&
+ \left[ 1 - 2\psi - \left( \partial_k \partial_l \Psi - \frac13 \delta_{kl} \partial^2 \Psi \right) x^k x^l \right] \frac{\delta_{ij} dx^i dx^j}{\left(1+K_F\, r^2/4\right)^2} \Bigg\}\,.
\label{eq:hFws}
\ea
Note that $\phi,\,\psi$ are defined with respect to $a_F$ and are thus 
distinguished from the usual parameterisation in global coordinates.  The crucial point is however
that the CFC frame is directly related to observations (see Sec.~6 of \cite{Dai:2015rda}), so
that this form of the metric is actually a desirable feature.  
For the time being, $\phi$ and $\psi$ are allowed to be fully nonlinear.

We now want to solve for the evolution of $\phi,\,\psi$ in the presence
of the long-wavelength mode $\Phi,\,\Psi$.  At first, one would expect
to have to solve the \emph{nonlinear Einstein equations}, that is, at second
and higher order in cosmological perturbation theory.  However, as described more generally in Sec.~5 of \cite{Dai:2015rda}, one of the
key virtues of the CFC framework is that this is actually not necessary, if one restricts to the regime in which small-scale perturbations have long entered the horizon (while the wavelength of the long mode is not restricted).  Note that for adiabatic perturbations, there are no interesting dynamics at all until the short-wavelength modes enter the horizon, and the long-wavelength mode, being far outside the horizon at this point, cannot have any dynamical impact.  

The reasoning proceeds as follows.  First of all, 
$a_F$ and $K_F$ can simply be incorporated into the background part of the
Einstein tensor. Further, for $k_S \gg \coH$ we can identify $\psi=\phi$, and only use the {\it scalar} Einstein equations, i.e.~the $00$-component as well as the divergence of the $0i$-component.  Finally, we only need to solve the Einstein equations on the central geodesic $G$, i.e.~at
$x^i = 0$, since we are free to choose the desired fluid trajectory
of interest as central geodesic, and since we do not need to take additional
spatial derivatives of the Einstein equations.  Clearly, only terms in the nonlinear
Einstein tensor with two spatial derivatives acting on the $\Phi$ or $\Psi$
pieces can contribute to the long-short coupling contribution to the Einstein equations on $G$.  Since the
Einstein tensor is linear in second derivatives, the short-wavelength perturbation in these terms can %at second order
only contain $\phi,\,\psi$ without any derivatives (these terms come from contracting indices with the perturbed metric).  In other
words, the only second-order terms in the Einstein equations would have
to be of the form
\ba
F(\phi) \left( \partial_i \partial_j \Psi - \frac13 \delta_{ij} \partial^2 \Psi \right)\,,
\ea
and correspondingly with $\Psi\to\Phi,\,\phi\to\psi$.  However, in order for these terms to contribute to scalar Einstein equations, the
only tensor available to contract with is $\delta_{ij}$.  Hence we
see that these terms have to vanish.

We emphasize that even the case $k_S \leq \coH$ is computable in CFC (as demonstrated in Sec.~7 of \cite{Dai:2015rda} for the case of long-wavelength tensor perturbations), although the calculation becomes technically involved because the trace-free part of the $ij$-Einstein equation must be used and terms of second order in perturbations start to appear in the Einstein tensor. Given the limited phenomenological significance of the regime $k_S \leq \coH$, we will focus on $k_S \gg \coH$ in the remainder of this section.  This implies that we can set $\phi=\psi$ even at second order.

To summarize: the \emph{isotropic} part of long-wavelength metric perturbations
is entirely absorbed into the background $a_F(\tau_F)$ and $K_F$, and thus
does not need to be dealt with explicitly; while the \emph{anisotropic} part
does not enter the Einstein equations for symmetry reasons.  Thus, 
in order to solve for the long-short mode coupling in the CFC
frame, \emph{it is sufficient to solve the purely small-scale Einstein equations with a modified scale factor}.  This holds as long as one can restrict to subhorizon short modes which are being influenced by long modes of arbitrary scale.  
The same was found to 
be true in \cite{Dai:2015rda} for long-wavelength tensor instead of scalar modes.  

In order to close the system, we need to also consider the fluid equations
\be
\nabla_\mu T^{\mu}{}_{\nu} = 0\,.
\ee
This is again very simple, since we merely have to include a modified
background and a tidal force which preserves its ``Newtonian'' form
on all scales [\refeq{hFws}].  Again, this is closely related to the
analogous problem of long-wavelength tensor modes studied in \cite{Dai:2015rda}.

%%%%%%%%%%%%%%%%%%%%%%%%%%%%%%%%%%%%%%%%%%%%%%%%%%%%%%%%%%%%%%%%%%%%%%%%%%%
\subsection{Outline of the calculation}

Let us outline the calculation that will take up the remainder of 
\refsec{app-squeezed-matter-bispec}.  We will consider the
leading coupling of long- and short-wavelength modes at second order 
in perturbation theory, 
so that we arrive at a second order small-scale solution
\be
\phi^{(2)}(\v{k}_L+\v{k}_S) \sim \Phi^{(1)}(\v{k}_L) \phi^{(1)}(\v{k}_S)\,,
\label{eq:phi2}
\ee
where ${}^{(1)}$ indicates linear order solutions.  We emphasize that this
is merely for simplicity. The CFC construction works for both nonlinear large-scale
and small-scale metric perturbations, and the simplification described above is not restricted
to the leading order only.  This is one of the key advantages of this approach.  
A higher order calculation in terms of the small-scale fluctuations has
recently been performed using N-body (``separate universe'') simulations
\cite{Wagner:2014aka,Li:2014jra,Wagner:2015gva}.  Going to higher order in the long-wavelength fluctuations
is beyond the scope of this paper and will be pursued elsewhere.  
As discussed in detail in Sec.~5 of \cite{Dai:2015rda}, it is then sufficient
to solve the linear Einstein equations for the long mode, and insert the
solution for the long-mode metric perturbation into the small-scale fluid
equations.  

We will begin by calculating the evolution of small-scale modes in the presence of long-wavelength perturbations, i.e.~the long-short mode coupling measurable by a CFC observer, separately for the cases of isotropic and anisotropic long-wavelength perturbations (\refsec{5.iso}--\refsec{5.vel}).  We then consider the mapping of CFC coordinates between neighboring observers (\refsec{5.eulerian}).  This leads to a final expression for the locally measurable squeezed matter bispectrum in \refsec{5.sqlocal}, to be understood in a sense described there.

The results of \refsec{5.iso} to \refsec{5.sqlocal} will be derived for a flat Einstein-de Sitter (EdS) background.  This is merely for calculational simplicity.  By replacing the EdS scale factor $a(t)$ with the linear growth factor $D(t)$, these results are numerically highly accurate even for other cosmologies, e.g. $\Lambda$CDM.

We conclude this outline by presenting the relevant equations.  
Since the long-wavelength density perturbation is
entirely absorbed into the local background $a_F(\tau_F)$, the only long-mode contribution to be included in the fluid equations is the local velocity field 
induced by the long-wavelength mode, which we
denote as $v^i_L$.  In CFC [\refeq{hFws}], the equation of motion for the latter becomes
\be
v^{(1)}_{L,i}{}' + \coH_F\, v^{(1)}_{L,i} + K_{ij}^\Phi\, x^j = 0\,.
\label{eq:vLeom}
\ee
This can be integrated to give
\be
v^{(1)}_{L,i}(\tau) =  -\frac{x^j}{a_F(\tau)}\int_0^\tau a_F(\tau')  K^\Phi_{ij}(\tau')  d\tau' \, = - \frac{2 x^j}{3\coH_F} \left(\partial_i \partial_j - \frac13 \delta_{ij} \partial^2\right) \Phi_{\rm ini},
\ee
where we have defined the long-wavelength tidal tensor along $G$ as
\be
K^{\Phi}_{ij} = \left[\partial_i \partial_j - \frac13 \delta_{ij} \partial^2\right] \Phi(\v{0},\tau_F)\,,
\ee
and the second equality is specific to an EdS universe.

The 00-Einstein equation, continuity equation, and Euler equation for
the small-scale modes then become
\bea
\partial^2 \phi & = & \frac32 \coH_F^2 \Om^F\, \delta \vs
\delta' + \partial_i[v^i + (1+\delta) v_L^i] & = & 0  \vs
v_i ' + \coH_F\, v_i + (v_L \cdot \partial) v_i + (v \cdot \partial) v_{L,i} & = & - \partial_i \phi \,,
\label{eq:Eulers1}
\eea
where all spatial derivatives are with respect to $x^i = x_F^i$.  
Here and throughout, $v^i$ denotes the small-scale velocity field, i.e.~the
total velocity in CFC without the tidal contribution $v^i_L$ due to the long-wavelength mode.  Note that we have dropped terms purely quadratic in small-scale perturbations in 
\refeq{Eulers1} in keeping with our second-order treatment in this paper.
 
\refeq{Eulers1} are almost identical to the well known system of equations used to solve
the evolution of large-scale structure on subhorizon scales.  The only
impact of the long-wavelength mode, apart from the modified cosmology
entering via $\coH_F,\,\Om^F$, is by adding the tidal-field induced velocity
$v^i_L$.  
Note that we have to keep this velocity to order $x^i$ in \refeq{Eulers1}  since
the evolution of the density field on $G$ depends on derivatives of $v_L^i$ on $G$.  
For an EdS background, the local matter density parameter in CFC is given by
\be
\Om^F - 1 = \frac{K_F}{a^2_F H_F^2} = \frac{K_F}{\coH_F^2}\,,
\label{eq:OmF}
\ee
where $K_F$ is the local curvature parameter as discussed in \refsec{local-friedmann-eqs}.  
In \refeq{Eulers1} we have neglected terms of the form $K_F \phi$ and replaced the curved-space
Laplacian $\gamma^{ij} \nabla_i \nabla_j$ with the flat-space Laplacian.  
If $K_F$ was of order $\coH^2_F$, these terms would be suppressed on subhorizon scales
with respect to the remaining terms by the same order of magnitude
as terms such as $\coH_F^2\, \phi,\,\coH_F\, \phi'$ which we have neglected.  
In fact, $K_F/\coH_F^2 \sim \Delta_{\rm sc} \ll 1$ by assumption, 
and so these are even smaller.  Current observations
constrain the curvature of the observable universe to be $|K/H_0^2| \lesssim 0.01$~\cite{Planck:2015xua}.  
Thus, the curvature terms in \refeq{Eulers1} are highly suppressed for
any practical subhorizon calculation.

We solve \refeq{Eulers1} as follows. We begin with linear solutions for the short scale variables $\delta$, $v$ and $\phi$, i.e.~neglecting $v^i_L$.  We then correct short-wavelength scalar perturbations due to their coupling with long-wavelength modes. With this procedure, the solution, taking $\delta$ for example, is 
\be
\delta = \delta^{(1)} + \delta^{(2)}\,,
\ee
where $\delta^{(1)}$ is the linear solution,  $\delta^{(2)}$ is the 
second-order correction due to mode coupling, and correspondingly for $\phi$, $v^i$ and the velocity divergence. In this context it is simplest to work with the proper velocity divergence, $\hat\theta \equiv a_F^{-1} \partial_i v^i$ which takes into account the local scale factor.    
Further, we will switch from $\tau_F$ to $t_F$ in the following.  This is convenient, since eventually we would like to compare $\delta$ with $\delta^{(1)}$ in the fiducial cosmology at fixed proper time $t_F$.  
 
The growing mode for the potential is constant, i.e.~$\phi=\phi_{\rm ini}$. The linear (subhorizon) solution reads
\ba
\label{eq:eq-5.10-linear-sol}
\delta^{(1)} = \frac{2}{3\coH^2} \partial^2 \phi_{\rm ini}, \quad v^{(1)}_i = - \frac{2}{3\coH} \partial_i \phi_{\rm ini}, \quad \hat\theta^{(1)} = - \frac{2}{3a \coH} \partial^2 \phi_{\rm ini}\,,
\ea
which obey equations for the {\it fiducial} cosmology with global expansion rate $\coH$ and zero curvature $\Omega_m=1$. The continuity and Euler equations for the second order small-scale
perturbations [i.e.~those of the order given in \refeq{phi2}] then become
\bea
\dot\delta^{(2)} + \hat \theta^{(2)} & = & - a^{-1}_F v_L^i \,\partial_i \delta^{(1)} \nn\\
%FS: l.h.s. is incorrect
%\dot {\hat \theta}^{(2)} + 2 H_F \hat \theta^{(2)} + \frac32 \Om^F H_F^2 \delta^{(2)} & = & - a^{-1}_F v_L^i \,\partial_i \hat\theta^{(1)} - 2 a^{-2}_F (\partial_i v_L^j)(\partial_j v_{(1)}^i)\nn\\
\dot {\hat \theta}^{(2)} + 2 H \hat \theta^{(2)} + \frac32 H^2 \delta^{(2)} & = & - a^{-1}_F v_L^i \,\partial_i \hat\theta^{(1)} - 2 a^{-2}_F (\partial_i v_L^j)(\partial_j v_{(1)}^i)\nn\\
&& - 2 \left( H_F - H \right) \hat \theta^{(1)} - \frac32 \left( \Omega^F_m H^2_F - H^2 \right) \delta^{(1)}\,,
\label{eq:Euler2nd}
\eea
where dots denote derivatives with respect to proper time $t_F$.  In the first line, we have used the fact that $v^i_L$ is divergence-free due to its purely tidal nature.  In \refeq{Euler2nd}, all coefficient functions
%, e.g. $\coH_F(\tau_F(t_F))$ versus $\coH(\tau(t))$, 
should be evaluated at fixed proper time (which does not imply fixed conformal time as the scale factor is also modified).  That is, $H_F(t_F)$ is the Hubble rate as function of time in the local modified cosmology, while $H(t_F)=2/(3t_F)$ is the same relation in the fiducial background.  

In order to solve for $\delta^{(2)}$ and $\hat\theta^{(2)}$, we need initial
conditions at early times $\tau_F \to 0$.  In this paper, we will set
the initial conditions to zero, $\delta^{(2)}|_{\rm ini} = \hat\theta^{(2)}|_{\rm ini} = 0$,
corresponding to no long-short mode coupling in the initial conditions.    
Ref.~\cite{Pajer:2013ana} showed that in single-field inflation in the
attractor regime, there exists no local-type non-Gaussianity in the CFC
frame, which implies that $\delta^{(2)}_{\rm ini}$ can only contain
long-mode contributions of order $k_L^2 \Phi|_{\rm ini}$.  In the
scale-invariant case, they scale as 
$(k_S/\coH)^2\, [(k_L/k_S)^2\, \Phi(\bfk_L) \phi(\bfk_S)|_{\rm ini}]$.  
These contributions,
which are typically phrased in terms of ``equilateral'' and ``orthogonal''
bispectrum shapes,
come from physical interactions during inflation, and are enhanced for example
for an inflaton with small sound speed.  A calculation of the proper
second order initial conditions including such terms is beyond the
scope of this paper and will be considered in future work.  
Neglecting these terms is justified for ``vanilla'' single-field inflation
with $c_s \sim 1$, since these contributions are of order
$(k_L/\coH)^2\, \Phi(\bfk_L) \phi(\bfk_S)$ and thus suppressed by
$(\coH/k_S)^2$ (which eventually saturates at $(\coH/k_{\rm eq})^2  $) relative to the terms we are keeping.

%%%%%%%%%%%%%%%%%%%%%%%%%%%%%%%%%%%%%%%%%%%%%%%%%%%%%%%%%%%%%%%%
\subsection{Initial conditions}
\label{sec:5.IC}
%%%%%%%%%%%%%%%%%%%%%%%%%%%%%%%%%%%%%%%%%%%%%%%%%%%%%%%%%%%%%%%%  

In order to solve for $\delta^{(2)}$ and $\hat\theta^{(2)}$, we need initial
conditions at early times $\tau_F \to 0$.  We are interested in deriving
the late-time matter statistics for initial conditions provided by
single-clock inflation.  Thus, we need the small-scale density contrast
in CFC in the presence of a long-wavelength mode, i.e. at second order,
predicted by single-clock inflation.  We obtain this by transforming
the second-order result in the form of the bispectrum of the curvature
perturbation $\mathcal{R}$ in comoving gauge as presented in \cite{Maldacena:2002vr}
to CFC.  This calculation has been presented in detail in \cite{Pajer:2013ana}, and we here summarize the results in order to make this section self-contained.  

In the notation of \cite{Maldacena:2002vr}, but replacing $\zeta_{\rm there}$ with $\mathcal{R}$ in order to be consistent with our convention,\footnote{Note that \cite{Pajer:2013ana} followed \cite{Maldacena:2002vr} and used the symbol $\zeta$.} the long-wavelength metric perturbations in comoving gauge are given by 
\ba
g_{00} =\:& a^2 [-1 -2 N_1] \vs
g_{0i} =\:& a^2\, N_i \vs
g_{ij} =\:& a^2 \big[ (1+ 2\mathcal{R}) \delta_{ij} + \gamma_{ij}\big]\,,
\label{eq:gcomoving}
\ea
where $\gamma_{ij}$ is tranverse-traceless and contains the tensor perturbations, which we will neglect in the following (see \cite{Dai:2015rda} for a CFC treatment of couplings between long-wavelength tensor modes and short-scale density modes~\cite{Dai:2013kra,Schmidt:2013gwa}). Note that we work to linear order here as we are only dealing with the long-wavelength mode, making use again of the advantages of the CFC formalism when dealing with a separation of scales.  
For the attractor solution of single-field inflation, the constraint equations
in this gauge yield \cite{Maldacena:2002vr}
\be
N_1 = \frac{\dot{\mathcal{R}}}{H} = \frac{\mathcal{R}'}{aH} \propto \frac{k_L^2}{(aH)^2}\,.
\label{eq:N1}
\ee
In this paper, we will ignore contributions from the initial conditions that scale as $k_L^2$, as they contain physical effects due to a nontrivial inflaton sound speed and thus depend on the particular inflationary model considered.  We will discuss this momentarily.

Neglecting $N_1$ following the discussion above, the tetrad corresponding to the metric \refeq{gcomoving} is given by
\ba
(e_0)^\mu =\:& a^{-1} (1, - N^i) \vs
(e_k)^\mu =\:& a^{-1} (1, [1 - \mathcal{R}] \delta_k^{\  i}) \,.
\ea
Note that this implies that the proper time interval is $dt_F = a(\tau) d\tau$.    It is now clear that for a perturbation $\mathcal{R}$ constant in space and time (as both are suppressed by powers of $k_L$), the only effect is to rescale the scale factor by a \emph{constant} amount.  The leading contributions in the small-$k_L$ limit to the local Hubble rate are of order $k_L^2 \mathcal{R}$.  This suppression of both spatial and time derivatives makes physical sense:  if $k_L \sim \coH$, then the local Hubble patch corresponds to a curved vacuum energy-dominated FLRW universe with curvature $(k_L/\coH)^2 \mathcal{R}$.  As the universe continues to inflate, with $H \approx$~const, the curvature decays as $a^{-2}$ and local Hubble patch rapidly asymptotes to flat de Sitter space (disregarding other perturbations that have been generated in the meantime).  

Given the $H_F = H$ to the order in $k_L$ that we work in, and our freedom to rescale $a_F$ by a constant, we can simply choose $a_F = a$ and thus $\tau_F = \tau$.  
At lowest order in derivatives of the metric \refeq{gcomoving}, the transformation to CFC at fixed $\tau=\tau_F$ is given by $x^\mu = (e_i) x_F^i$, that is
\ba
x^i =\:& (1-\mathcal{R}) x_F^i\,.
\ea
Note that at this order in derivatives, the shift vector $N_i$ drops out.  
Of course, we could have equally absorbed $\mathcal{R}$ into the definition of $a_F$, yielding the same end result.

The small-scale curvature perturbations $\mathfrak{r}$ (to be distinguished from the long-wavelength perturbation $\mathcal{R}$) then simply transform into CFC via
\be
\mathfrak{r}_{\rm CFC}(\tau,\bfx_F) = \left[1 - \mathcal{R}\, x_F^i\partial_i\right] \mathfrak{r}_{\rm global}(\tau,\bfx_F)\,.
\ee
This implies that the power spectrum of $\mathfrak{r}$ transforms as
\be
P_{\mathfrak{r}}^{\rm CFC}(k_S)\Big|_{\mathcal{R}} = P_{\mathfrak{r}}^{\rm global}(k_S)
\left[1 + \mathcal{R} \frac{d\ln k_S^3 P_{\mathfrak{r}}(k_S)}{d\ln k_S}\right]
= \left[1 + (n_s-1) \mathcal{R}\right] P_{\mathfrak{r}}^{\rm global}(k_S)
\,,
\ee
where $n_s-1$ is the slope of the dimensionless power spectrum
$k^3 P_{\mathfrak{r}}$ evaluated at $k_S$, and $P_{\mathfrak{r}}$ without
superscript denotes the expectation value of the small-scale power spectrum
(marginalized over long-wavelength modes).  As shown in \cite{Maldacena:2002vr},
the squeezed-limit bispectrum of curvature perturbations obeys the famous
relation
\be
B^{\rm global}_{\mathcal{R}}(k_L, k_S) = - (n_s-1) P_{\mathcal{R}}(k_L) P_{\mathfrak{r}}(k_S) + \mathcal{O}\left(\frac{k_L^2}{k_S^2}\right)\,,
\ee
which is equivalent to a local modulation of the small-scale power
spectrum by
\be
P_{\mathfrak{r}}^{\rm global}(k_S)\Big|_{\mathcal{R}}
= \left[1 + (n_s-1) \mathcal{R}\right] P_{\mathfrak{r}}^{\rm global}(k_S)\,.
\ee
This modulation is \emph{exactly canceled} by the transformation to CFC,
i.e.
\be
P_{\mathfrak{r}}^{\rm CFC}(k_S)\Big|_{\mathcal{R}} = P_{\mathfrak{r}}(k_S) + \mathcal{O}(k_L^2 \mathcal{R})\,.
\ee
That is, a local observer measuring the power spectrum of small-scale curvature perturbations during single-clock inflation \emph{cannot detect the presence of a long-wavelength mode up to order} $\mathcal{O}(k_L^2 \mathcal{R})$.  

Thus, for the calculation of small-scale perturbations in CFC frame,
we have to set the initial conditions of the long-short mode coupling to
zero, $\mathfrak{r}^{(2)} = 0$.  Correspondingly, 
$\delta^{(2)}|_{\rm ini} = \hat\theta^{(2)}|_{\rm ini} = 0$.  We emphasize
again that for this we only needed to rely on a correct calculation of
the leading long-short mode coupling (equivalently, the squeezed-limit
bispectrum) in \emph{some global coordinates}, along with the \emph{linear}
relation of the metric perturbation in these coordinates to the late-time
density field.  We do not need to rely on a second-order relation of
$\mathcal{R}$ or $\mathfrak{r}$ with a nonlinear definition of spatial
curvature on some preferred spatial slice, as discussed in \cite{Bruni:2013qta,Bruni:2014xma,Bertacca:2015mca}.

Let us briefly discuss the corrections of order $k_L^2 \mathcal{R} \propto k_L^2 \Phi$ that we neglect here.  In the scale-invariant case, they scale as 
$(k_S/\coH)^2\, [(k_L/k_S)^2\, \Phi(\bfk_L) \phi(\bfk_S)|_{\rm ini}]$.  
These contributions,
which are typically phrased in terms of ``equilateral'' and ``orthogonal''
bispectrum shapes,
come from physical interactions during inflation, and are enhanced for example
for an inflaton with small sound speed.  A calculation of the proper
second order initial conditions including such terms is beyond the
scope of this paper and will be considered in future work.  
Neglecting these terms is justified for ``vanilla'' single-field inflation
with $c_s \sim 1$, since these contributions are of order
$(k_L/\coH)^2\, \Phi(\bfk_L) \phi(\bfk_S)$ and thus suppressed by
$(\coH/k_S)^2$ (which eventually saturates at $(\coH/k_{\rm eq})^2  $) relative to the terms we are keeping.

%%%%%%%%%%%%%%%%%%%%%%%%%%%%%%%%%%%%%%%%%%%%%%%%%%%%%%%%%%%%%%%%
\subsection{Isotropic case: local expansion and curvature}
\label{sec:5.iso}
%%%%%%%%%%%%%%%%%%%%%%%%%%%%%%%%%%%%%%%%%%%%%%%%%%%%%%%%%%%%%%%%  

We first consider an isotropic long-wavelength perturbation, which implies
$K^\Phi_{ij} =0$ and thus $v_{L,i}=0$.  Restricting \refeq{Euler2nd} to this case, taking a time derivative of the continuity equation and combining with the Euler equation, we derive one second order equation for $\delta^{(2)}$,
\bea
\label{eq:eq-delta2-iso-fixedtF}
\ddot\delta^{(2)} + 2 H(t_F) \dot\delta^{(2)} - \frac32 H^2(t_F) \delta^{(2)} & = & - 2 \left[ H_F(t_F) - H(t_F) \right] \dot\delta^{(1)} \nn\\
&& + \frac32 \left[ H^2_F(t_F) \Omega^F_m(t_F) - H^2(t_F) \right] \delta^{(1)},
\eea 
There are two approaches to relate the terms on the r.h.s. to the long-wavelength perturbations.  One can either go back to the expressions
in global coordinates [\refeq{local-Hubble} and \refeq{KF-comoving-curvature}].  This is detailed in \refapp{HFOmF}.  Alternatively, one can use the fact that the scale factor $a_F(t_F)$ obeys
the Friedmann equation for a curved, matter-dominated universe, resulting
in a Hubble rate given by
\be
H_F^2(a_F) = H_{F0}^2 \left[\Omn^F a_F^{-3} + \Omega_{K0}^F a_F^{-2} \right]
\ee
where $H_{F0} = H_F(a_F=1)$ and $\Omega_{K0}^F = -K_F/H_{F0}^2$.  
Note that $H_F^2 \Om^F a_F^{3}$ is a constant,
proportional to the comoving matter density.  Evaluating this at early time,
where $H_F \to H, \Om^F \to 1, a_F(t_F) \to a(t_F)$ [\refeq{aFIC}], we obtain (evaluated at fixed $t_F$)
\be
H_F^2 \Om^F a_F^{3} = H^2 a^{3} \quad\Rightarrow\quad
H_F^2(t_F) \Om^F(t_F) = H^2(t_F) \frac{a^3(t_F)}{a_F^3(t_F)} \,.
\label{eq:HFOmF}
\ee
Now, $H_F^2 \Om^F$ is the physical matter density in the
CFC region (up to a constant $3/(8\pi G)$), while $H^2$ is proportional to the
fiducial background matter density \emph{at the same proper time}.  
Since the ratio between those matter densities is the density
perturbation in synchronous gauge, we have
\be
\frac{a^3(\tau_F)}{a_F^3(\tau_F)} = 1 + \Delta_{\rm sc}(\tau_F)\,.
\label{eq:aFDelta}
\ee
Note that this agrees with the r.h.s. of \refeq{aFovera} after the difference between CFC time $dt_F=a_F(\tau_F)d\tau_F$ and global time $dt=a(\tau)d\tau$ is accounted for.  We then obtain
\be
H_F^2(t_F) \Om^F(t_F) - H^2(t_F) = H^2(t_F) \Delta_{\rm sc}(t_F)
= \frac{2}{3a^2} \partial^2 \Phi\,,
\ee
where we have used \refeq{Poissonsc} in the Einstein-de Sitter background
($\partial^2 \Phi = 3\coH^2 \Delta_{\rm sc}/2$).  
Note that this in fact holds to nonlinear order in $\Delta_{\rm sc}$
\cite{Wagner:2014aka}.  

Now, going back to \refeq{HFOmF}, we can evaluate this when $a_F = 1$,
which on the l.h.s. yields $H_{F0}^2 \Omn^F$.  We can also evaluate it at $a=1$, 
for which the r.h.s. yields $H_0^2$, which proves that $H_{F0}^2 \Omn^F = H_0^2$.  
Thus, at linear order,
\ba
H_F(t_F) - H(t_F) =\:& \frac1{2H(t_F)}\left[H_F^2 - H^2\right]_{t_F} 
= \frac1{2H(t_F)}\left[ H_0^2 a_F^{-3} - K_F a^{-2} - H_0^2 a^{-3} \right]_{t_F}
\vs
=\:& \frac12 H(t_F) \left[ \Delta_{\rm sc} - \frac{K_F}{H_0^2} a \right]_{t_F}
= -\frac29 \frac{H(t_F)}{\coH^2(t_F)} \partial^2 \Phi\,.
\ea
With these results, \refeq{eq-delta2-iso-fixedtF} reduces to
\bea
\ddot \delta^{(2)} + 2 H \dot \delta^{(2)} - \frac32 H^2 \delta^{(2)} & = & \left( \frac{13}{9} \frac{1}{a^2} \partial^2 \Phi \right) \delta^{(1)}\,.
\eea
With the initial condition $\delta^{(2)}|_{t_F=0}=0$ and $\dot \delta^{(2)}|_{t_F=0}=0$ imposed, we find the modified small-scale density field in the presence of the long-wavelength mode
\bea
\label{eq:delta-2nd-isotropic}
\delta^{(2)} & = & \left( \frac{26}{63} \frac{1}{\coH^2} \partial^2 \Phi \right) \delta^{(1)} = \frac{13}{21}\,\Delta_{\rm sc}\,\delta^{(1)}\,,
\eea
where all functions are evaluated at the same proper time.  
The second equality again uses \refeq{Poissonsc}.  
\refeq{delta-2nd-isotropic} agrees with solving spherical collapse in a homogeneous, but curved FLRW universe \cite{Baldauf:2011bh,Wagner:2015gva}, once the overdensity of that universe has been identified with the density perturbation in synchronous-comoving gauge.

%%%%%%%%%%%%%%%%%%%%%%%%%%%%%%%%%%%%%%%%%%%%%%%%%%%%%%%%%%%%%%%%
\subsection{Anisotropic case: tidal field}
\label{sec:5.aniso}
%%%%%%%%%%%%%%%%%%%%%%%%%%%%%%%%%%%%%%%%%%%%%%%%%%%%%%%%%%%%%%%% 

Following our discussion, we now set the isotropic contribution to
zero, i.e.~$H_F \to H$ and $\Om^F \to \Om=1$ in \refeq{Euler2nd}, and
keep only the terms involving $v^i_L$.  
We obtain the differential equations
\ba
\dot\delta^{(2)} + \hat\theta^{(2)} =\:& - a^{-1}\, v_L^i \,\partial_i \delta^{(1)} \vs
\dot{\hat\theta}^{(2)} + 2 H\, \hat\theta^{(2)} + \frac32 H^2 \delta^{(2)}
=\:& - a^{-1} v_L^i \,\partial_i \hat\theta^{(1)} - 2 a^{-2} \left( \partial^i v_L^j \right) \left( \partial_j v^{(1)}_i \right)\,.
\label{eq:EuleraSH}
\ea
An equivalent system of equations (within Newtonian gauge on small scales) was solved for EdS in \cite{Schmidt:2013gwa}.  Using their results [Eqs.~(38), (39) and (45) there], we obtain
\bea
\label{eq:delta-2nd-anisotropic}
\delta^{(2)} & = & \frac4{7}\,K^\Delta_{ij}\, \left( \frac{\partial^i\partial^j}{\partial^2} \delta^{(1)} \right),
\eea
where the anisotropic density tensor is defined as
\bea
K^\Delta_{ij} & = & \frac2{3\coH^2} K^\Phi_{ij} = \left( \frac{\partial_i \partial_j}{\partial^2} - \frac13 \delta_{ij} \right) \Delta_{\rm sc}\,.
\eea

%%%%%%%%%%%%%%%%%%%%%%%%%%%%%%%%%%%%%%%%%%%%%%%%%%%%%%%%%%%%%%%%
\subsection{Velocity divergence}
\label{sec:5.vel}
%%%%%%%%%%%%%%%%%%%%%%%%%%%%%%%%%%%%%%%%%%%%%%%%%%%%%%%%%%%%%%%%  

We would also like to derive the velocity divergence $\hat\theta$ in CFC at fixed proper time $t_F$.  For this, notice that the continuity equation in \refeq{Euler2nd}, restricted to $x^i=0$ so that $v_L^i=0$, reads
\bea
\hat\theta & = & - \dot\delta\,.
\eea
We can thus simply take the time derivative
of \refeq{d2sum} to obtain the second-order correction to $\hat\theta$ due to long-wavelength modes on the geodesic,
\bea
\hat\theta^{(2)} = H \left[\frac{26}{21}\,\Delta_{\rm sc}\,\delta^{(1)} + \frac8{7}\,K^\Delta_{ij}\, \left( \frac{\partial^i\partial^j}{\partial^2} \delta^{(1)} \right)\right]\,.
\label{eq:hattheta2sum}
\eea
The commonly used velocity divergence $\theta$ defined with respect to comoving coordinates at this order is then simply $\theta^{(2)} = a\, \hat\theta^{(2)}$.  

The result for velocity divergence is useful because for pure potential flow (valid in the regime of perfect fluid) the velocity shear tensor can be derived from it. The velocity shear tensor is needed in order to derive the effects of redshift-space distortions on clustering statistics.

%%%%%%%%%%%%%%%%%%%%%%%%%%%%%%%%%%%%%%%%%%%%%%%%%%%%%%%%%%%%%%%%
\subsection{Quasi-local Eulerian fields}
\label{sec:5.eulerian}
%%%%%%%%%%%%%%%%%%%%%%%%%%%%%%%%%%%%%%%%%%%%%%%%%%%%%%%%%%%%%%%%  

Combining \refeq{delta-2nd-isotropic} and \refeq{delta-2nd-anisotropic}, we write down the second-order small-scale density field quantifying the long-short mode coupling, strictly {\it on the central geodesic},
\bea
\delta_G & = & \left( 1 + \frac{13}{21} \Delta_{\rm sc} \right) \delta^{(1)} + \frac4{7}\, K^\Delta_{ij} \, \left( \frac{\partial^i\partial^j}{\partial^2} \delta^{(1)} \right).
\label{eq:d2sum}
\eea
This describes the modified clustering induced by a long-wavelength gravitational potential that a local observer would measure along $G$ relative to the average small-scale modes at fixed proper time. However, in order to measure clustering observables such as power spectra and bispectra, the observer also needs to survey how density varies across a finite region surrounding $G$. A computational trick is to choose the CFC central geodesic to pass through a given spacetime point under examination, and compute local physical quantities within that particular CFC. It is shown in Sec.~6 of \cite{Dai:2015rda} that measurements from neighboring CFC observers can be related by adding the local geodesic deviation.

To be more specific, neighboring CFC observers are Lagrangian observers, since the coordinate system is tied to the fluid trajectory.  Their measurements of the density field on each individual geodesic can be collected and mapped to a density field across a quasi-local Eulerian coordinate patch via
\bea
\delta_E & = & \delta_G - s^i\, \partial_i \delta_G, 
\label{eq:deltaEG}
\eea 
where the geodesic deviation vector $s^i$ is given by~\footnote{One does not have to distinguish between $a_F$ and $a$ here because $h^F_{00}$ is first-order in the long mode.}
\bea
s^i = \frac12\,q^j \int^{\tau_F}_0 \frac{d\tau'}{a_F(\tau')} \int^{\tau'}_0 d\tau'' a_F(\tau'') \left( \partial_i \partial_j h^F_{00} \right),
\eea
where $h^F_{00} = - K^\Phi_{ij} x^i x^j$, and the Lagrangian coordinate $q^j$ can be identified with $x^j$ at the perturbative order we seek. For Einstein-de Sitter universe we find
\bea
s_i & = & - \frac{2}{3 \coH^2}\, K^\Phi_{ij}\, x^j = - K^\Delta_{ij}\, x^j.
\eea 
We will refer to the second term in \refeq{deltaEG} as the ``displacement'' contribution.  
The quasi-local Eulerian density field then reads
\bea
\label{eq:eq-5.44-deltaE-res}
\delta_E & = & \delta^{(1)} + \frac{13}{21} \Delta_{\rm sc}\, \delta^{(1)} + \frac4{7}\, K^\Delta_{ij} \, \left( \frac{\partial^i\partial^j}{\partial^2} \delta^{(1)} \right)  + K^\Delta_{ij}\,x^i \partial^j \delta^{(1)}\,.
\eea
In the same way, the velocity divergence (with respect to comoving coordinates) in quasi-local Eulerian frame, including the ``displacement'' term, reads
\bea
\label{eq:eq-5.38-thetaE}
\theta_E & = & \theta^{(1)} - \frac{26}{21} \coH\, \Delta_{\rm sc} \, \delta^{(1)} - \frac{8}{7} \coH\,K^\Delta_{ij} \left( \frac{\partial^i \partial^{j}}{\partial^2} \delta^{(1)} \right)  + K^\Delta_{ij}\,x^i \partial^j \theta^{(1)}\,.
\eea 
Note that the explicit dependence on $x^i$ in the displacement terms will enter in the power spectrum as a derivative with respect to scale via the Fourier transform.   
\refeqs{eq-5.44-deltaE-res}{eq-5.38-thetaE} are very similar to the expressions for the second-order density and velocity divergence in Eulerian Standard Perturbation Theory (SPT), which are restricted to subhorizon scales for all modes.  
Here we have generalized these results to a large-scale mode of arbitrary wavelength, albeit in the limit $k_L \ll k_S$.  The key differences to the SPT result are that first, the long-wavelength mode $\Delta_{\rm sc},\,K^\Delta_{ij}$ is in \emph{synchronous-comoving gauge} (since $k_S \gg \coH$, the gauge choice for the small-scale modes is not important).  Second, the displacement term only enters as a \emph{relative} displacement $\propto x^j$, which removes the gauge artifact in the unphysical absolute displacement present in the SPT expression. 

\refeqs{eq-5.44-deltaE-res}{eq-5.38-thetaE}
thus clearly show that intrinsic contributions to matter clustering due to long-short mode-coupling, which generate gravitational non-Gaussianity, are proportional to $\Delta_{\rm sc}$ or $K^\Delta_{ij}$, both of which scale as $\partial^2 \Phi / \coH^2$ rather than the gravitational potential $\Phi$ itself. This holds for long-wavelength modes on arbitrarily large scales.  This has to be so, as the {\it Equivalence Principle} demands that 
constant or pure-gradient metric perturbations do not produce locally observable gravitational effects.  
Thus, any coupling to small-scale modes of the long-wavelength potential $\Phi(\v{k}_L)$ itself has to be imprinted in the initial conditions. This does not happen in single-clock inflation.

The expressions in \refeqs{eq-5.44-deltaE-res}{eq-5.38-thetaE}, %{eq-sigmaE} 
representing locally measurable fields, do show some differences from the SPT results in terms of the numerical coefficients.  This is due to two effects.  
First, within the local CFC patch, density fluctuations have been normalized with respect to the ``local mean density''
\bea
\rho_F(t_F) & = & \bar\rho(t_F) \left[ 1 + \Delta_{\rm sc}(t_F) \right],
\eea
which is modified by the long mode. Normalizing fluctuations with respect to the global $\bar\rho$ would result in an additional correction $\Delta_{\rm sc}\delta^{(1)}$ to the second-order density. Secondly, for both density and velocity gradient, the comoving coordinates (CFC) have been defined with respect to the local scale factor
\bea
a_F(t_F) & = & a(t_F) \left[ 1 - \Delta_{\rm sc}(t_F)/3 \right].
\eea
Switching back to the global scale factor $a$ amounts to providing an additional ``displacement'' term, which is equivalent to adding back the trace part to $K^\Delta_{ij}$ in the displacement $s^i$.  
A detailed description of how to relate our results \refeqs{eq-5.44-deltaE-res}{eq-5.38-thetaE} to SPT expressions can be found in \refapp{CFC-SPT}.  
Note that both of these additional corrections only involve $\Delta_{\rm sc} \propto \partial^2\Phi/\coH^2$ and obey the Equivalence Principle.  

%%%%%%%%%%%%%%%%%%%%%%%%%%%%%%%%%%%%%%%%%%%%%%%%%%%%%%%%%%%%%%%%
\subsection{Squeezed matter bispectrum as measured by a local observer}
\label{sec:5.sqlocal}
%%%%%%%%%%%%%%%%%%%%%%%%%%%%%%%%%%%%%%%%%%%%%%%%%%%%%%%%%%%%%%%%  

Following \refeq{eq-5.44-deltaE-res}, the CFC observer measures a local power spectrum of short-scale density fluctuations (in the presence of long-wavelength perturbation),
\bea
\left. P_\delta(\bfk_S) \right|_{\Delta(\bfk_L)}  & = & \left[ 1 + K^\Delta_{ij}\,\hat k^i_S \hat k^j_S \left( \frac87 - \frac{d\ln P_\delta(k_S)}{d\ln k_S} \right) + \frac{26}{21} \Delta_{\rm sc} \right] P_\delta(k_S),
\label{eq:Pklocal}
\eea
where $P_\delta(k)$ is the linear power spectrum of density fluctuations (in cN gauge, although this is not relevant at the scale $k_S$).  
\refeq{Pklocal} can then be used to calculate a ``locally observable squeezed-limit matter bispectrum'' as follows.  Consider a class of observers distributed over large scales which measure the local mean density $\rho_F$ and small-scale power spectrum [l.h.s. of \refeq{Pklocal}] in their environment at fixed proper time.  They then communicate this information to a distant observer in the future light cone of the entire class of observers.  This distant observer in the future can then construct an estimate of the ensemble average of $\langle \Delta(\bfk_L) P_\delta(\bfk_S)|_{\Delta(\bfk_L)}\rangle$, which is equivalent to the squeezed matter bispectrum.  The ensemble average of what he would measure is then 
\bea
\label{eq:squeezed-local-matter-bispec}
\hspace{-1cm}
\VEV{ \delta(\bfk_1) \delta(\bfk_2) \Delta_{\rm sc}(\bfk_L) }' & = & \left[ \frac{26}{21} + \left( \mu^2_{SL} - \frac13 \right) \left( \frac87 - \frac{d\ln P_\delta(k_S)}{d\ln k_S} \right) \right] P^\Delta_{\rm sc}(k_L) P_\delta(k_S).
\eea
where $\bfk_S = \left( \bfk_1 - \bfk_2 \right)/2$ and $\mu_{SL} \equiv \hat \bfk_L \cdot \hat \bfk_S$ and we require $|\bfk_1| \approx | \bfk_2 | \approx k_S \gg k_L$. For an EdS universe, the linear matter power spectrum for the long mode in sc gauge is related to the primordial potential power spectrum via $P^\Delta_{\rm sc}(k_L) = (4/9\coH^4) k^4_L\, P_\Phi(k_L)$.  

It is worth stressing that the scaling behaviour of the squeezed-limit bispectrum 
would be dramatically different for primordial non-Gaussianity of the local type~\cite{Komatsu:2001rj} in the initial conditions,
\bea
\Phi_{\rm NG}\Big|_{\rm ini} & = & \Phi + f^{\rm loc}_{\rm NL} \left( \Phi^2 - \VEV{\Phi^2} \right),
\label{eq:PhiNGloc}
\eea
where the non-Gaussian potential $\Phi_{\rm NG}$ is a local function of a Gaussian potential $\Phi$. In that case, the density field at second order is expressed through the first-order Gaussian fields via
\bea
\delta^{(2)}\Big|_{\rm ini} & = & 2 f^{\rm loc}_{\rm NL}\,\Phi\,\delta^{(1)} \Big|_{\rm ini}\,.
\eea 
Instead of its second derivative, the gravitational potential $\Phi$ is directly involved, and would lead to a contribution $\propto f^{\rm loc}_{\rm NL}\, P^{\Phi \Delta_{\rm sc}}(k_L)$ in \refeq{squeezed-local-matter-bispec}, which leads to order unity departures from our result if $k_L/\coH \lesssim (f^{\rm loc}_{\rm NL})^{1/2}$.  
The same effect gives rise to a scale-dependent bias for matter tracers on large scales~\cite{Dalal:2007cu}. From our analysis within the CFC formalism, {\it we do not find an observable $f^{\rm loc}_{\rm NL}$ or a scale-dependent bias from general relativistic clustering, nor from a general relativistic specification of initial condition for single-clock inflation. Any locally observable effect of this type has to be imprinted in the initial conditions by a departure from single-clock inflation.} This is different from the conclusions in Refs.~\cite{Bartolo:2005xa,Verde:2009hy,Bruni:2013qta,Bertacca:2015mca}. In combination with the results of \cite{Pajer:2013ana}, this implies that in single-field inflation in the attractor regime, $f^{\rm loc}_{\rm NL}-$type contributions to the local matter bispectrum and halo bias vanish identically;  in particular, there is no $f^{\rm loc}_{\rm NL} \sim (n_s-1)$ contribution to observables. Further, this constraint also applies to local-type non-Gaussianity of higher order, where higher powers of $\Phi$ are added to \refeq{PhiNGloc}, starting with $g_{\rm NL}^{\rm loc} \Phi^3$.

Of course, \refeq{squeezed-local-matter-bispec} does not correspond to how the matter bispectrum is measured in practice. In experiments, astrophysical tracers of the matter distribution are charted according to their observed redshift and their apparent position in the sky. Therefore, their number density, or the underlying matter density, appears distorted to a distant observer, since photons emitted from the source region are subject to change in the frequency and the direction of propagation as they travel from the source to the observer. Relevant effects include boundary effects such as redshift-space distortion and the Sachs-Wolfe effect, as well as line-of-sight integrated effects such as gravitational lensing, the Integrated Sachs-Wolfe effect, and time delay. In a full general relativistic treatment, individual effects are gauge dependent, only the sum of them is gauge invariant. Detailed discussions of the projection effects are beyond the scope of this paper. We refer interested readers to a systematic study of the relativistic projection effects on the galaxy power spectrum~\cite{Jeong:2013psa} and on the squeezed galaxy bispectrum~\cite{Scoccimarro:1999ed,Pajer:2013ana,Kehagias:2015tda}.

Projection effects can mimic the $f^{\rm loc}_{\rm NL}$ parameter in the apparent clustering.  {\it However, they arise purely as a result of photon propagation, and are physically unrelated to the nonlinear gravitational dynamics that occur locally} (general relativistic or not). Projection effects in general are sensistive to the late-time expansion history as well as properties of source objects such as luminosity function and bias. Thus, by observing tracers with different bias and luminosity functions, it is in principle possible to measure the projection contributions separately and disentangle them from the local properties of the matter density field which we have derived here.

%%%%%%%%%%%%%%%%%%%%%%%%%%%%%%%%%%%%%%%%%%%%%%%%%%%%%%%%%%%%%%%%%%%%%%%%%%%%%%%
%%%%%%%%%%%%%%%%%%%%%%%%%%%%%%%%%%%%%%%%%%%%%%%%%%%%%%%%%%%%%%%%%%%%%%%%%%%%%%%
\section{Summary and conclusions}
\label{sec:concl}
%%%%%%%%%%%%%%%%%%%%%%%%%%%%%%%%%%%%%%%%%%%%%%%%%%%%%%%%%%%%%%%%%%%%%%%%%%%%%%%
%%%%%%%%%%%%%%%%%%%%%%%%%%%%%%%%%%%%%%%%%%%%%%%%%%%%%%%%%%%%%%%%%%%%%%%%%%%%%%%

In this work, we have proved that an inertial observer in a flat FLRW universe perturbed by a long-wavelength density perturbation (isotropic in the sense that $\partial_i\partial_j \Phi = \delta_{ij} \partial^2\Phi/3$) would locally experience a modified FLRW expansion, with a renormalized scale factor $a_F$, expansion rate $H_F$ and an induced effective spatial curvature $K_F$. Those results have been rigorously derived by explicit construction of the CFC around observer's geodesic.  Since $a_F$ and $K_F$ are uniquely determined as local observables (up to a multiplicative constant), there is no ambiguity in this construction.

The necessary conditions for this ``separate universe'' picture are that the anisotropic stress vanishes and that sound horizons of all fluids are vanishing or negligibly small compared to the scale of the long-wavelength density fluctuation. We have presented expressions for $H_F$ and $K_F$, at linear order in the long-wavelength perturbation amplitude and valid on arbitrarily large scales. The former includes a contribution from metric perturbations, in addition to peculiar velocity divergence. The latter is directly proportional to the Laplacian of the gauge-invariant curvature perturbation on comoving slices $\mathcal{R}$. The CFC construction guarantees that these results are gauge-independent.  
Using this fact, the impact of long-wavelength density perturbations on small-scale dynamics can be numerically simulated by implementing the modified expansion rate $a_F$ and the modified spatial curvature $K_F$~\cite{Wagner:2014aka}.   
This technique has a wide range of applications. For instance, the reponse of tracer abundance to a change in the ``background'' cosmology is related to tracer bias~\cite{Baldauf:2011bh}. Another example involves power-spectrum response functions measured comparing separate universes with different FLRW parameters, which give the angle-averaged squeezed matter higher-point correlations~\cite{Wagner:2015gva}.  

Generalizing to \textit{anisotropic} long-wavelength perturbations, the leading deviation~\footnote{Corrections of higher order in the spatial departure from the observer's trajectory would generate observable effects on small-scale dynamics that are suppressed by powers of the scale ratio $k_L/k_S$. These effects will have phenomenological relevance beyond the squeezed limit.} from FLRW cosmology is a trace-free tidal contribution to the metric that is quadratic in the spatial CFC coordinates.  The tidal tensor is exactly in the Newtonian form $(\partial_i \partial_j - \delta_{ij} \partial^2/3 ) \Phi$ without general relativistic corrections, if the scalar potential in conformal-Newtonian gauge $\Phi$ is used. Since by construction CFC eliminates (essentially) all gauge freedom, the tidal tensor is also independent of the gauge one chooses to parameterize long-wavelength perturbations. Implementation of the tidal tensor into separate universe simulations could be a potential technique to study tidal alignment of halo shape or galaxy shape.

For an isotropic configuration of the long-wavelength perturbation, Birkhoff's theorem guarantees that the ``separate universe'' can be generalized to all orders in the amplitude of the long mode along the observer's geodesic (as proved in \refsec{proof}, for the tractable case of a compensated tophat). For the more general anisotropic case, the local expansion history $a_F(t_F)$ is expected to violate the Friedmann equation at second order, as suggested by the presence of an extra term, quadratic in the anisotropic velocity shear, in the Raychaudhuri equation~\cite{Raychaudhuri:1955cosmology}. Despite the breakdown of the ``separate universe'' conjecture, our construction of CFC will remain applicable. In that case, it would depict a \emph{locally} uniformly and isotropically expanding spacetime, again up to tidal metric corrections.

Another major result of this work is the \textit{squeezed-limit matter bispectrum} measured by a local observer [\refeq{squeezed-local-matter-bispec}] for an EdS cosmology, which we derive using CFC without having to resort to the full second-order Einstein equations. While for simplicity the subhorizon limit is assumed for the short-scale modes, the full general relativistic dynamics of the long mode are accounted for by the CFC formalism. Long-wavelength perturbations generate a squeezed bispectrum through their influence on the local gravitational clustering.  This contribution takes the same form as in the familiar SPT result within Newtonian gravity, as long as the long-wavelength density contrast in sc gauge is used. In other words, there are no explicit corrections due to general relativity. This curious aspect at second order is analogous to the observation at linear order that the Newtonian Poisson equation can be extended to arbitrary scales if the sc-gauge density contrast and the cN-gauge potential are used~\cite{Chisari:2011iq}.  In order to relate this bispectrum to observations from large-scale structure surveys, projection effects due to photon propagation through a perturbed background need to be included.  While the ruler perturbations of~\cite{Schmidt:2012ne} in principle provide all the necessary projection effects, we leave the detailed calculation of these to future work.

%%%%%%%%%%%%%%%%%%%%%%%%%%%%%%%%%%%%%%%%%%%%%%%%%%%%%%%%%%%%%%%%
\acknowledgments

F.S. would like to thank Sam Ip and Christian Wagner for helpful discussions.
L.D. was supported by the John Templeton Foundation as a graduate research associate. E.P. is supported by the Delta-ITP consortium, a program of the Netherlands organization for scientific research (NWO) that is funded by the Dutch Ministry of Education, Culture and Science (OCW). Tensorial algebra are partially performed with {\tt xPand}~\cite{Pitrou:2013hga}.

%%%%%%%%%%%%%%%%%%%%%%%%%%%%%%%%%%%%%%%%%%%%%%%%%%%%%%%%%%%%%%%%%%%%%%%%%%%%%%%
%%%%%%%%%%%%%%%%%%%%%%%%%%%%%%%%%%%%%%%%%%%%%%%%%%%%%%%%%%%%%%%%%%%%%%%%%%%%%%%
%\clearpage
\appendix
%%%%%%%%%%%%%%%%%%%%%%%%%%%%%%%%%%%%%%%%%%%%%%%%%%%%%%%%%%%%%%%%%%%%%%%%%%%%%%%
%%%%%%%%%%%%%%%%%%%%%%%%%%%%%%%%%%%%%%%%%%%%%%%%%%%%%%%%%%%%%%%%%%%%%%%%%%%%%%%
%
%%%%%%%%%%%%%%%%%%%%%%%%%%%%%%%%%%%%%%%%%%%%%%%%%%%%%%%%%%%%%%%%%%%%%%%%%%%%%%
\section{Spatial gauge freedom}
\label{app:spatialgauge}
%%%%%%%%%%%%%%%%%%%%%%%%%%%%%%%%%%%%%%%%%%%%%%%%%%%%%%%%%%%%%%%%%%%%%%%%%%%

In the presence of long-wavelength scalar perturbations, the space-space components of the CFC tidal metric (\refeq{eq-189}) can be written as
\bea
\label{eq:eq-108}
&& h^F_{ij} =  \frac13 K_F \left( x_i x_j - \delta_{ij} r^2 \right) - \delta_{ij} \left( \partial_k \partial_l \Psi - \frac13 \delta_{kl} \partial^2 \Psi \right) x^k x^l \\
&& + \left[ \frac13 \left( \delta_{li} \partial_j \partial_k + \delta_{lj} \partial_i \partial_k + 2 \delta_{ij} \partial_k \partial_l - \delta_{kl} \partial_i \partial_j \right) \Psi - \frac19 \left( \delta_{ik} \delta_{jl} + \delta_{il} \delta_{jk} + \delta_{ij} \delta_{kl} \right) \partial^2 \Psi \right] x^k x^l. \nn
\eea
In the first line, the first term is proportional to the local spatial curvature. The second term, proportional to $\delta_{ij}$, is equal to the same tidal potential in $h_{00}$ if $\Psi=\Phi$. This is the expected term satisfying the conformal Newtonian gauge condition $h_{ij} = \delta_{ij}\, h_{00}$ in the absence of anisotropic stress. Below we first show that the curvature term is equivalent to the stereographic form \refeq{stereographic-curved-FLRW} for a curved FLRW universe. We then show that the second line can be eliminated by a re-parameterization of the spatial coordinates, and therefore has no physical consequences.

We use the following spatial diffeomorphism: up to including $\mathcal{O}[(x^i)^2]$, the most generic re-parameterization of spatial coordinates which leaves $h_{00}$ and $h_{0i}$ invariant reads \cite{Dai:2015rda}
\bea
\label{eq:spatial-diffeo}
x^i \longrightarrow x^i + \frac16 A^i{}_{jkl}(\tau)\, x^j x^k x^l.
\eea
The rank-4 tensor $A^i{}_{jkl}$ is fully symmetric with respect to its last three indices. It is allowed to be a function of $\tau$, and the time-dependence only affects $h_{00}$ and $h_{0i}$ at $\mathcal{O}[(x^i)^3]$. The change to the line element is
\bea
dx^i = dx^i + \frac12 A^i{}_{jkl}\, x^k x^l\, dx^j,
\eea
which changes $h_{ij}$ by
\bea
\label{eq:eq-111}
h_{ij} \longrightarrow h_{ij} + \frac12 \left( A_{i,jkl} + A_{j,ikl} \right) \, x^k x^l.
\eea

%%%%%%%%%%%%%%%%%%%%%%%%%%%%%%%%%%%%%%%%%%%%%%%%%%%%%%%%%%%%%%%%
\subsection{The effective curvature}
\label{sec:LS-effective-curvature}
%%%%%%%%%%%%%%%%%%%%%%%%%%%%%%%%%%%%%%%%%%%%%%%%%%%%%%%%%%%%%%%% 

We have demonstrated in \refsec{local-friedmann-eqs} that long-wavelength scalar perturbations must induce an effective curvature $K_F$ in the local-coordinate description. Here we see that the effective curvature is residing in the first term of \refeq{eq-108}.

Now in \refeq{spatial-diffeo} we choose
\bea
A^i{}_{jkl} & = & - \frac16 K_F \left( \delta^i{}_j \delta_{kl} + \delta^i{}_k \delta_{jl} + \delta^i{}_l \delta_{jk} \right),
\eea
which generates radial reparameterization \refeq{eq-106}. This changes that metric term by
\bea
\frac13 K_F \left( x_i x_j - \delta_{ij} r^2 \right) & \longrightarrow & - \frac12 K_F \delta_{ij} r^2,
\eea
and exactly accounts for the lowest order curvature contribution of the stereographic parameterization \refeq{stereographic-curved-FLRW}.

It can be concluded that as long as $K_F$ is conserved, the only local {\it isotropic} effect of long-wavelength scalar perturbations, at $\mathcal{O}[(x^i_F)^2]$, is an effective FLRW curvature, with no further general relativistic correction.

%%%%%%%%%%%%%%%%%%%%%%%%%%%%%%%%%%%%%%%%%%%%%%%%%%%%%%%%%%%%%%%%
\subsection{Conformal Newtonian coordinates for CFC}
\label{app:cN-CFNC}
%%%%%%%%%%%%%%%%%%%%%%%%%%%%%%%%%%%%%%%%%%%%%%%%%%%%%%%%%%%%%%%%

What about the anisotropic tidal terms? Since we only have rank-2 tensors $\partial_i \partial_j \Psi$, $\partial^2 \Psi$ and $\delta_{ij}$ at our disposal, the most generic $A_{i,jkl}$ that has the right symmetry can be written as
\bea
A_{i,jkl} & = & a \left( \delta_{ij} \delta_{kl} + \delta_{ik} \delta_{jl} + \delta_{il} \delta_{jk} \right) \partial^2 \Psi + b \left( \delta_{ij} \partial_k \partial_l + \delta_{il} \partial_k \partial_j + \delta_{ik} \partial_l \partial_j \right) \Psi \nn\\
&& + c \left( \delta_{kl} \partial_i \partial_j + \delta_{jl} \partial_i \partial_k + \delta_{jk} \partial_i \partial_l \right)\Psi,
\eea
where $a$, $b$ and $c$ are arbitrary numerical coefficients. By matching the coefficients of various terms, we find a unique solution in order for the change in \refeq{eq-111} to exactly compensate the second line of \refeq{eq-108}, $a = - 1/9, \,b = 2/3, \,c = - 1/3$. With a successive change of coordinates in \refeq{eq-106}, we arrive at the following simple form for the CFC metric
\bea
ds^2 & = & a^2_F(\tau_F) \left[ - \left( 1 + \left( \partial_k \partial_l \Phi - \frac13 \delta_{kl} \partial^2 \Phi \right) x^k x^l \right) d\tau^2_F \right.\nn\\
&& \left. + \left( 1 - \left( \partial_k \partial_l \Psi - \frac13 \delta_{kl} \partial^2 \Psi \right) x^k x^l \right) \frac{\delta_{ij} dx^i dx^j}{\left(1+K_F\, r^2/4\right)^2} \right].
\eea
We therefore have put the CFC metric into the conformal Newtonian form, which reduces exactly to the Newtonian coordinates on sub-horizon scales. It is manifest in this conformal Newtonian frame that all the anisotropic effects of the long-wavelength perturbation are parameterized by a simple tidal potential term.

%%%%%%%%%%%%%%%%%%%%%%%%%%%%%%%%%%%%%%%%%%%%%%%%%%%%%%%%%%%%%%%%
\section{CFC expansion rate and matter density}
\label{app:HFOmF}
%%%%%%%%%%%%%%%%%%%%%%%%%%%%%%%%%%%%%%%%%%%%%%%%%%%%%%%%%%%%%%%%

Applying \refeq{local-Hubble} and \refeq{KF-comoving-curvature} to EdS universe with growing perturbations, we have
\bea
\label{eq:eq-5.18}
H_F(t_F) - H(t) & = & - H \left( \Phi + \frac{2}{9\coH^2} \partial^2 \Phi \right), \\
\label{eq:eq-5.19}
H^2_F(t_F)\, \Omega^F_m(t_F) - H^2(t) & = & \frac{K_F}{a^2_F} + H^2_F(t_F) - H^2(t) = - 2 H^2 \Phi + \frac{2}{3 a^2} \partial^2 \Phi.
\eea
The global time $t$ is understood to refer to {\it the same spacetime point} which has proper time $t_F$ in CFC. Therefore, $t$ and $t_F$ are numerically different, which can only be neglected on the right hand sides of the equations which explicitly contain first-order perturbation variables. We furthermore need to account for modification of the proper time due to long-wavelength perturbations, and derive $H(t) - H(t_F) = - \dot H\,\Delta t$, with $\Delta t = t_F - t =2\Phi/3H$, and similarly for $H^2(t)-H^2(t_F)$. We eventually find
\bea
H_F(t_F) - H(t_F) & = & - \frac{2H}{9\coH^2} \partial^2 \Phi, \\
H^2_F(t_F)\, \Omega^F_m(t_F) - H^2(t_F) & = & \frac{2}{3 a^2} \partial^2 \Phi.
\eea
Compared with \refeqs{eq-5.18}{eq-5.19}, we see that all terms proportional to the Newtonian potential $\Phi$ itself cancel out; locally observable effects must depend on second spatial derivatives $\partial^2\Phi$. 

%%%%%%%%%%%%%%%%%%%%%%%%%%%%%%%%%%%%%%%%%%%%%%%%%%%%%%%%%%%%%%%%
\section{The 2nd Friedmann equation}
\label{app:Fr2}
%%%%%%%%%%%%%%%%%%%%%%%%%%%%%%%%%%%%%%%%%%%%%%%%%%%%%%%%%%%%%%%%  

For completeness, we should also verify that the second Friedmann equation holds, which can be done in an analogous way. It follows from \refeq{eq-CFNC-acceleration} that the acceleration is given by,
\bea
\label{eq:eq-acceleration}
\frac{1}{a^2_F} \frac{d \coH_F}{d\tau_F} = \frac{1}{a_F} \frac{d^2 a_F}{dt^2_F} = \frac{1}{a} \frac{d^2 a}{dt^2} - \frac{1}{a^2} \left[ \coH \Phi' + 2 \coH' \Phi + \coH \Psi' + \Psi'' - \frac13 \left( \partial \cdot V' + \coH \partial \cdot V \right) \right],
\eea
where $dt_F=a_F(\tau_F) d\tau_F$ is the CFC proper time. We now use the trace (isotropic) part of the space-space Einstein equation (note that there is in general a total pressure perturbation on the right-hand side),
\bea
2 \Psi'' + \coH \left( 4 \Psi' + 2 \Phi' \right) + 2 \left( 2 \coH' + \coH^2 \right) \Phi = 8 \pi G a^2 \left( \mathcal{P}_F - \bar{\mathcal{P}} \right),
\eea
We then have
\bea
&& \coH \Phi' + 2 \coH' \Phi + \coH \Psi' + \Psi'' -  \frac13 \left( \partial\cdot V' + \coH \partial\cdot V \right) \nn\\
& = & \frac13 \left[ \partial^2 \Psi - 3 \coH \left( \Psi' + \coH \Phi \right) \right] + 4\pi G a^2 \left( \mathcal{P}_F - \bar{\mathcal{P}} \right) + \frac13 \partial^2 \left(\Phi - \Psi \right) \nn\\
& = & \frac{4\pi G}{3} a^2 \left[ \left( \rho_F - \bar\rho \right) + 3 \left( \mathcal{P}_F - \bar{\mathcal{P}} \right) \right] + \frac13 \partial^2 \left(\Phi - \Psi \right),
\eea
via the time-time Einstein equation. Again, the second term vanishes if there is no anisotropic stress. Inserting this into \refeq{eq-acceleration}, we immediately obtain the form of the second Friedmann equation,
\bea
\label{eq:eq-2nd-Friedmann-CFNC}
\frac{1}{a_F} \frac{d^2 a_F}{dt^2_F} = - \frac{4 \pi G}{3} \left( \rho_F + 3 \mathcal{P}_F \right),
\eea
where again $\rho_F$ and $\mathcal{P}_F$ are respectively the total energy density and pressure at the observer's location. Contrary to the discussion of the first Friedmann equation, no assumption on the cosmic fluids is needed for the second Friedmann equation to hold.

%%%%%%%%%%%%%%%%%%%%%%%%%%%%%%%%%%%%%%%%%%%%%%%%%%%%%%%%%%%%%%%%
\section{Relating CFC result to SPT result}
\label{app:CFC-SPT}
%%%%%%%%%%%%%%%%%%%%%%%%%%%%%%%%%%%%%%%%%%%%%%%%%%%%%%%%%%%%%%%% 

How does \refeq{eq-5.44-deltaE-res} compare to the usual result of Standard Perturbation Theory (SPT) at second order?  First of all, the density contrast measured in CFC is computed with reference to the ``local homogeneous density'', i.e.~$\rho_F = \bar\rho (1+\Delta_{\rm sc})$, while in SPT one makes reference to the global homogeneous density $\bar\rho$. This amounts to adding back a contribution $\Delta_{\rm sc}\,\delta^{(1)}$. Secondly, the ``displacement'' term in \refeq{eq-5.44-deltaE-res} is proportional to the trace-free $K^\Phi_{ij}$. This is because the trace part, corresponding to an isotropic convergent or divergent flow, has been absorbed into the local scale factor $a_F$. To undo this renormalization, one adds back a term of this isotropic flow (to find $a_F(t_F)/a(t_F)-1$ one uses \refeq{aFovera} and then accounts for $\Delta t = t_F-t$),
\bea
\left( \frac{a(t_F)}{a_F(t_F)} - 1 \right) x^i \partial_i \delta^{(1)} & = & \frac13 \Delta_{\rm sc}\,x^i \partial_i \delta^{(1)}.
\eea
Implementing these two corrections, we obtain the second-order density constrast
\bea
\delta^{(2)} & = & \left( \frac{\partial_i \partial_j}{\partial^2} \Delta_{\rm sc}  \right)\, x^j \partial_i \delta^{(1)} + \frac{10}{7} \Delta_{\rm sc}\,\delta^{(1)} + \frac47 \left( \frac{\partial_i \partial_j}{\partial^2} \Delta_{\rm sc} \right) \left( \frac{\partial^i \partial^j}{\partial^2} \delta^{(1)} \right).
\eea
Finally, the spatial origin is still chosen at the center of the local patch, and is therefore moving with it. One can undo the bulk motion of the local patch by including a homogeneous ``displacement'' (which leaves correlation functions invariant), and then adjust to whatever new spatial origin that is desired,
\bea
\left( \frac{\partial_i \partial_j}{\partial^2} \Delta_{\rm sc}  \right)\, x^j \partial_i \delta^{(1)} & \rightarrow & \left[ 1 + x^j \partial_j \right] \left( \frac{\partial_i}{\partial^2} \Delta_{\rm sc}  \right)\, \partial_i \delta^{(1)} \rightarrow \left( \frac{\partial_i}{\partial^2} \Delta_{\rm sc}  \right)\, \partial_i \delta^{(1)}.
\eea
We obtain
\bea
\delta^{(2)} & = & \left( \frac{\partial_i}{\partial^2} \Delta_{\rm sc}  \right)\, \partial_i \delta^{(1)} + \frac{10}{7} \Delta_{\rm sc}\,\delta^{(1)} + \frac47 \left( \frac{\partial_i \partial_j}{\partial^2} \Delta_{\rm sc} \right) \left( \frac{\partial^i \partial^j}{\partial^2} \delta^{(1)} \right).
\eea
These tensorial structures and numerical coefficients agree completely with the familiar $F_2$ kernel in second-order SPT computed for EdS universe~\cite{Bernardeau:2001qr}, in the limit that one wave number is much larger than the other, except for an overall factor of two, which accounts for swapping the long-wavelength and short-wavelength modes. We stress that the $F_2$ kernel is derived from Newtonian nonlinear clustering on subhorizon scales, while our result is still valid for (super-)horizon scale $k_L$. The Newtonian form is, however, preserved when the long-wavelength density contrast in sc gauge is used.

Performing a similar check on the velocity divergence \refeq{eq-5.38-thetaE}, we find agreement with the $G_2$ kernel of SPT.

%%%%%%%%%%%%%%%%%%%%%%%%%%%%%%%%%%%%%%%%%%%%%%%%%%%%%%%%%%%

%%%%%%%%%%%%%%%%%%%%%%%%%%%%%%%%%%%%%%%%%%%%%%%%%%%%%%%%%%%

\end{document}